\documentclass[12pt,12pt, draftclsnofoot, onecolumn]{IEEEtran}
\usepackage[latin9]{inputenc}
\usepackage{geometry}
\usepackage{array}
\usepackage{float}
\usepackage{color}
\usepackage{units}
\usepackage{mathtools}
\usepackage{mathrsfs}
\usepackage{multirow}
\usepackage{amsmath}
\usepackage{amssymb}
\usepackage{graphicx}
\usepackage{setspace}
\makeatletter

\providecommand{\tabularnewline}{\\}
\floatstyle{ruled}
\newfloat{algorithm}{tbp}{loa}
\providecommand{\algorithmname}{Algorithm}
\floatname{algorithm}{\protect\algorithmname}

\usepackage{amsfonts}
\usepackage{amsthm}
\usepackage{mathbbol}
\usepackage{algorithm}
\usepackage{algorithmic}
\usepackage{bbm}
\usepackage{cite}
\usepackage{fancyhdr}
\usepackage{latexsym}
\usepackage{multirow}
\usepackage{mathrsfs}
\usepackage{setspace}
\usepackage{stfloats}
\usepackage{subfigure}
\usepackage{url}
\usepackage{latexsym}
\theoremstyle{plain}
\ifCLASSINFOpdf
\else
\fi
\renewcommand{\maketag@@@}[1]{\hbox{\m@th\normalsize\normalfont#1}}%

\makeatother

\begin{document}
\title{Novel Visible Light Communication Assisted Perspective-Four-Line Algorithm
for Indoor Localization}
\author{Lin~Bai
\thanks{L. Bai is with the Beijing Key Laboratory of
Network System Architecture and Convergence, School of Information
and Communication Engineering, Beijing University of Posts and Telecommunications,
Beijing 100876, China (e-mail: bailin2126@bupt.edu.cn).}}
%\author{Lin~Bai, Yang~Yang,~\IEEEmembership{Member,~IEEE,} Chunyan~Feng,~\IEEEmembership{Senior~Member,~IEEE,}
%and~Caili~Guo,~\IEEEmembership{Senior~Member,~IEEE}% <-this % stops a space
%\thanks{L. Bai, Y. Yang and C. Feng are with the Beijing Key Laboratory of
%Network System Architecture and Convergence, School of Information
%and Communication Engineering, Beijing University of Posts and Telecommunications,
%Beijing 100876, China (e-mail: bailin2126@bupt.edu.cn; young0607@bupt.edu.cn;
%cyfeng@bupt.edu.cn).}% <-this % stops a space
%\thanks{C. Guo is with Beijing Laboratory of Advanced Information Networks,
%School of Information and Communication Engineering, Beijing University
%of Posts and Telecommunications, Beijing 100876, China (e-mail: guocaili@bupt.edu.cn).}% <-this % stops a space
%\thanks{This work was supported by National Natural Science Foundation of
%China (61871047), National Natural Science Foundation of China (61901047),
%Beijing Natural Science Foundation (4204106) and China Postdoctoral
%Science Foundation (2018M641278).}}
\markboth{}{}
\maketitle
\begin{abstract}
In this paper, we propose a novel visible light communication (VLC)
assisted Perspective-four-Line algorithm (V-P4L) for practical indoor
localization. The basic idea of V-P4L is to joint VLC and computer
vision to achieve high accuracy regardless of LED height differences. In particular,
we first exploit the space-domain information to estimate the orientation
and coordinate of a single rectangular LED luminaire in the camera
coordinate system based on the plane and solid geometry. Then, based
on the time-domain information transmitted by VLC and the estimated
luminaire information, V-P4L can estimate the position and pose of
the camera by the single-view geometry theory and the linear least
square (LLS) method. To further mitigate the effect of height differences
among LEDs on localization accuracy, we then propose a correction
algorithm of V-P4L based on the LLS method and a simple optimization
method. Due to the combination of time- and space-domain information,
V-P4L only requires a single luminaire for localization without limitation
on the correspondences between the features and their projections.
Simulation results show that for V-P4L the position error is always
less than $15\,\mathrm{cm}$ and the orientation error is always less
than $3{^{\circ}}$ using popular indoor luminaires.
\end{abstract}


\IEEEpeerreviewmaketitle{}

\section{Introduction}

\label{sec:intro} \IEEEPARstart{A}{ccurate} indoor localization
is increasingly important due to the surging position based services
such as position tracking, navigation and robot movement control.
In this research field, visible light positioning (VLP) technologies
and computer vision based localization have the advantage of high
accuracy and low cost. Visible light positioning technologies exploit
visible light signals for determining the position of the receiver.
Visible light possesses strong directionality and low multipath interference,
and thus VLP can achieve high accuracy localization performance \cite{Pathak2015Visible,lim2015ubiquitous,yang2019relay}.
Additionally, VLP utilizes light-emitting diodes (LEDs) as transmitter.
Benefited from the increasing market share of LEDs, VLP has relatively
low cost on infrastructure \cite{Pathak2015Visible,yang2016enhanced}.
On the other hand, computer vision based localization relies on the
images of the reference features captured by cameras to estimate the
position and pose of cameras with high accuracy \cite{piasco2018survey,ben2014review}.
Cameras can provide an extensive amount of information at limited
power consumption, small size and reasonable cost \cite{ben2014review}.
Additionally, nowadays cameras are essential parts of smartphones,
which further corroborates the feasibility of vision based localization
technologies \cite{do2016depth}. Therefore, VLP and computer vision
based localization have been gained increasing attentions in recent
years \cite{do2016depth,ICRA2018Manhattan}.

Typical VLP algorithms include proximity \cite{Sertthin20106}, fingerprinting
\cite{Qiu2016Let}, time of arrival (TOA) \cite{wang2013TOA}, angle
of arrival (AOA) \cite{zhu2017ADOA}, received signal strength (RSS)
\cite{bai2019camera,bai2020enhanced,li2014epsilon,yasir2015rssjlt}
and image sensing \cite{li2018vlc}. Proximity and fingerprinting
cannot estimate the receiver pose even though only a single luminaire
is required. Additionally, the accuracy of proximity is insufficient
\cite{bai2019camera} while fingerprinting requires at least three
luminaires to reduce the effect of ambiguity issues \cite{vegni2012indoor}.
Among these VLP algorithms, RSS algorithms are most widely-used due
to their high accuracy and low cost \cite{bai2019camera}. However,
RSS algorithms require multiple luminaires for localization, like
image sensing, TOA and AOA algorithms \cite{do2016depth}. Moreover,
RSS algorithms rely on accurate channel model, which is challenging
to be achieved in practice. A popular assumption in RSS algorithms
is that the radiation pattern of LEDs is the Lambertian model which
may not be true for many luminaires especially when a lampshade is
used \cite{miramirkhani2015channel}. Meanwhile, the estimated channel
gain is affected by sunlight, dust and shadowing in practice \cite{Dong2017study,cailean2017current}.
Therefore, the feasibility of RSS algorithms is limited.

On the other hand, typical computer vision based localization methods
include Perspective-n-Line (PnL) and Perspective-n-Point (PnP). The
methods are usually performed by analyzing $n$ correspondences between
three dimensional (3D) reference features and their two dimensional
(2D) projections on the image (i.e., 3D-2D correspondences), where
the features are either points or lines \cite{xu2016pose}. In particular,
PnP methods employ the point features, while PnL methods employ the
line features. Compared with the point features, line features can
carry richer information \cite{xu2016pose}. Therefore, compared with
PnP methods, PnL methods can achieve higher detection accuracy and
are more robust to occlusions \cite{pvribyl2017absolute,vakhitov2016accurate}.
However, PnL methods need 3D-2D correspondences which are difficult
to obtain. In existing PnL studies, the 3D-2D correspondences assumed
to be perfectly known in advance \cite{xu2016pose,xiaojian2008analytic},
which is impractical in practice \cite{ICRA2018Manhattan}. To circumvent
this challenge, the work in \cite{ICRA2018Manhattan} proposed a method
to find the 3D-2D correspondences for the scenario where the number
of the vertical lines are more than that of horizontal lines. However,
this method cannot be applied to the scenario where there is no significant
difference between the numbers of horizontal and vertical lines, such
as the scenario where the rectangular beacons are deployed on the
ceiling. Therefore, the feasibility of the method is constrained.

The main contribution of this paper is a novel visible light communication
(VLC) assisted Perspective-four-Line algorithm (V-P4L), which can
achieve feasible and accurate indoor localization. \textit{To the
authors' best knowledge, this is the first localization algorithm
that only requires a single luminaire}\footnote{All the LEDs in the luminaire transmit the same information for ease
of implementation. Note that the proposed algorithm can also be implemented
when LEDs in the luminaire transmit different information. In this
case, the 3D-2D correspondence can be obtained directly from the different
information. However, this requires higher implementation complexity
and the robustness of the link may be affected by inter-channel-interference.
Therefore, in this work, we adopt the former strategy for higher robustness
and lower complexity.}\textit{ for position and pose estimation without given
3D-2D correspondence}. The key contributions of this paper include:
\begin{itemize}
\item We propose an indoor localization algorithm termed as V-P4L, which
uses camera to simultaneously capture the information in time and
space domains of LEDs to achieve high feasibility and high accuracy.
Based on the plane and solid geometry theory, V-P4L estimates the
luminaire's information in the camera coordinate system first using
the space-domain information of LEDs. Then, based on the single-view
geometry theory and the time-domain information of LEDs, V-P4L can
estimate the pose and position of the camera exploiting the luminaire's
information in different coordinate systems. In this way, V-P4L can
estimate the position and pose of the receiver only using a single
luminaire.
\item To avoid the requirement of the 3D-2D correspondences, the time-domain
information transmitted by VLC and the linear least square (LLS) method
are exploited in V-P4L to properly match 3D-2D correspondences. Based
on the time-domain information, V-P4L can obtain the information of
LEDs in the world coordinate. Then, based on the LLS method and the
LEDs' information in different coordinate systems, the 3D-2D correspondences
can be properly matched. In this way, V-P4L can achieve high feasibility.
\item To further improve the feasibility of V-P4L, we then propose a correction
algorithm for V-P4L to correct for the scenarios with LED height differences based on the information in both time and space domains. When LEDs have different heights, based on the single-view
geometry theory and the LLS method, the correction algorithm can first
estimate the pose and 2D position of the camera, and then based on
the single-view geometry theory and a simple optimization method,
the correction algorithm can estimate the 3D position of the camera.
In this way, V-P4L can be used regardless of the height differences
among LEDs.
\end{itemize}
Simulation results show that for V-P4L the position error is always
less than $15\,\mathrm{cm}$ and the orientation error is always less
than $3{^{\circ}}$ using popular indoor luminaires.

The rest of the paper is organized as follows. Section \ref{sec:System-Model}
introduces the system model. Section \ref{sec:luminaire information}
calculates the luminaire information in the camera coordinate system.
The proposed basic algorithm of V-P4L is detailed in Section \ref{sec:V-P4L-SH},
and the proposed correction algorithm is detailed in Section \ref{sec:V-P4L-DH}.
Simulation results are presented in Section \ref{sec:simulation}.
Finally, the paper is concluded in Section \ref{sec:CONCLUSION}.

The following notations are used throughout the paper: $A$ and $a$
with or without subscript denote scalars; $\mathbf{v}$ denotes a
column vector and $\mathbf{A}$ stands for a matrix; $\left|A\right|$
denotes the absolute value of $A$; $\mathbf{A}^{\mathrm{T}}$, $\mathbf{A}^{\mathrm{-1}}$,
$\det\left(\mathbf{A}\right)$ and $\left\Vert \mathbf{A}\right\Vert _{2}$
indicate the transpose, inverse, determinant and Eculidean norm of
$\mathbf{A}$, respectively; $\mathbf{v}\times\mathbf{u}$ denotes
the cross product of $\mathbf{v}$ and $\mathbf{u}$; $\mathbf{v}\cdot\mathbf{u}$,
$\mathbf{A}\cdot\mathbf{v}$ and $\mathbf{A}\mathbf{B}$ denotes the
dot products of $\mathbf{v}$ and $\mathbf{u}$, $\mathbf{A}$ and
$\mathbf{v}$, and $\mathbf{A}$ and $\mathbf{B}$, respectively;
$\hat{A}$, $\hat{\mathbf{A}}$ and $\hat{\mathbf{v}}$ represent
the estimate of $A$, $\mathbf{A}$ and $\mathbf{v}$, respectively.
In addition, there are some special or important symbols used throughout
in this paper, which are listed in Table \ref{tab:Symbols} with their
meaning. In particular, $P$ with subscript denotes the 3D vertex
of the LED luminaire; $p$ with subscript denotes the 2D point on
the image plane; $L$ with subscript represents the 3D line; $l$
with subscript represents the 2D line on the image plane. We use subscript
to represent the indices or objects that the scalars, points, vectors
and matrices corresponding to. For example, $P_{i}$ denotes the $i$th
vertex of the luminaire, and $\rho_{l_{ij}}$ denotes a parameter
in the equation of $l_{ij}$. Furthermore, we use the superscript
to represent the coordinates of points and vectors in different coordinate
systems. For example, the coordinates of the 2D point $p_{i}$ in
the world, camera, image and pixel coordinates are denoted as $p_{i}^{\mathrm{w}}$,
$p_{i}^{\mathrm{c}}$, $p_{i}^{\mathrm{i}}$ and $p_{i}^{\mathrm{p}}$,
respectively.

\global\long\def\arraystretch{1.1}%
\begin{table}
\caption{\label{tab:Symbols}Symbols and Their Meaning.}

\centering{}%
\begin{tabular}{r|>{\raggedright}p{12cm}}
\hline
Symbol & Meaning\tabularnewline
\hline
\hline
$\left(u_{0},v_{0}\right)^{\mathrm{T}}$ & Pixel coordinate of $o^{\textrm{i}}$\tabularnewline
\hline
$f$ & Focal length\tabularnewline
\hline
$f_{u}$, $f_{v}$ & Focal ratios\tabularnewline
\hline
$d_{x}$, $d_{y}$ & Physical size of each pixel\tabularnewline
\hline
$P_{i}$ & The $i$th vertex of the luminaire\tabularnewline
\hline
$p_{i}$ & Projection of $P_{i}$ on the image plane\tabularnewline
\hline
$p_{i}^{\textrm{p}}$/$p_{i}^{\textrm{i}}$ & Pixel/Image coordinate of $p_{i}$\tabularnewline
\hline
$P_{i}^{\textrm{c}}$/$P_{i}^{\textrm{w}}$ & Camera/World coordinate of $P_{i}$\tabularnewline
\hline
$L_{ij}$ & 3D reference line connecting $P_{i}$ and $P_{j}$\tabularnewline
\hline
$l_{ij}$ & 2D projection of $L_{ij}$ on the image plane\tabularnewline
\hline
$\mathbf{n}_{L_{ij}}^{\mathrm{c}}$ & Direction vector of $L_{ij}$ in CCS\tabularnewline
\hline
$\phi_{l_{ij}}$ & Rotate angle from $y^{\textrm{i}}$-axis to $l_{ij}$ in anticlockwise
direction\tabularnewline
\hline
$\rho_{l_{ij}}$ & Distance from $o^{\textrm{i}}$ to $l_{ij}$\tabularnewline
\hline
$\mathbf{n}_{\mathrm{LED}}^{\textrm{w}}$/$\mathbf{n}_{\mathrm{LED}}^{\textrm{c}}$ & Normal vector of the luminaire in WCS/CCS\tabularnewline
\hline
$\Pi_{ij}$ & Lateral face determined by the vertices $P_{i}$, $P_{j}$ and $o^{\textrm{c}}$\tabularnewline
\hline
$\mathbf{n}_{\Pi_{ij}}^{\mathrm{c}}$ & Normal vector of $\Pi_{ij}$ in CCS\tabularnewline
\hline
$S$ & Area of the luminaire\tabularnewline
\hline
$H$ & Distance from $o^{\textrm{c}}$ to the luminaire\tabularnewline
\hline
$V$ & Volume of rectangular pyramid $o^{\textrm{c}}-P_{1}P_{2}P_{3}P_{4}$\tabularnewline
\hline
$V_{i}$ & Volume of triangular pyramid $o^{\textrm{c}}-P_{i}P_{j}P_{k}$\tabularnewline
\hline
$\varphi$, $\theta$, $\psi$ & Euler angles corresponding to the $x^{\mathrm{c}}-$axis, $y^{\mathrm{c}}-$axis
and $z^{\mathrm{c}}-$axis\tabularnewline
\hline
$\mathbf{R}_{\mathrm{c}}^{\mathrm{w}}$/$\mathbf{t_{\mathrm{c}}^{\mathrm{w}}}$ & Pose/Position of the camera in WCS\tabularnewline
\hline
\end{tabular}
\end{table}
\vspace{-0cm}

\section{\label{sec:System-Model}System Model}

\begin{figure}[t]
\begin{centering}
\includegraphics[scale=0.6]{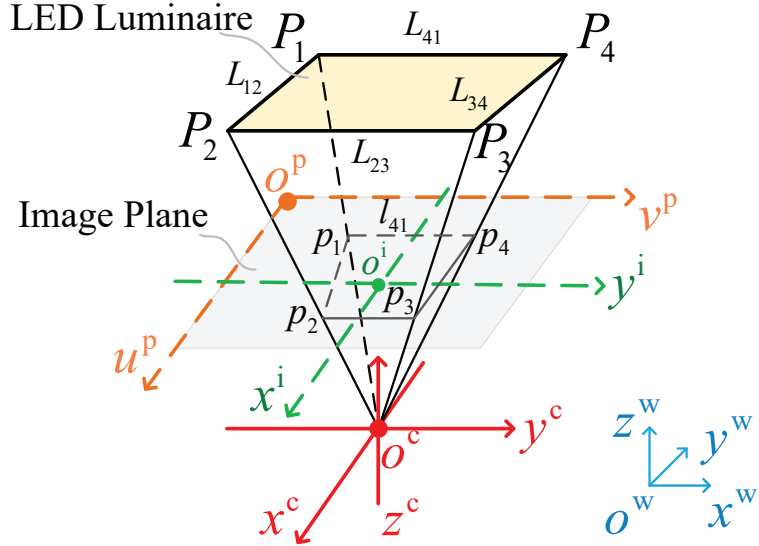}
\par\end{centering}
\caption{\label{fig:A-system-block}The system diagram of the proposed algorithm.}
\end{figure}

The system diagram is illustrated in Fig. \ref{fig:A-system-block}.
Four coordinate systems are utilized for localization, which are the
pixel coordinate system (PCS) $o^{\textrm{p}}-u^{\textrm{p}}v^{\textrm{p}}$
on the image plane\footnote{As shown in Fig. \ref{fig:A-system-block}, the image plane is a virtual
plane. In this paper, the camera is a standard pinhole camera. The
actual image plane is behind the camera optical center (i.e., the
pinhole), $o^{\textrm{c}}$. To show the geometric relations more
clearly, the virtual image plane is set up in front of $o^{\textrm{c}}$
as done in many papers \cite{masselli2014new,kneip2011novel}. In
particular, the virtual image plane and the actual image plane are
centrally symmetric, and $o^{\textrm{c}}$ is the center of symmetry.}, the image coordinate system (ICS) $o^{\textrm{i}}-x^{\textrm{i}}y^{\textrm{i}}$
on the image plane, the camera coordinate system (CCS) $o^{\textrm{c}}-x^{\textrm{c}}y^{\textrm{c}}z^{\textrm{c}}$
and the world coordinate system (WCS) $o^{\textrm{w}}-x^{\textrm{w}}y^{\textrm{w}}z^{\textrm{w}}$.
In PCS, ICS and CCS, the axes $u^{\textrm{p}}$, $x^{\textrm{i}}$
and $x^{\textrm{c}}$ are parallel to each other and, similarly, $v^{\textrm{p}}$,
$y^{\textrm{i}}$ and $y^{\textrm{c}}$ are also parallel to each
other. Additionally, $o^{\textrm{p}}$ is at the upper left corner
of the image plane and $o^{\textrm{i}}$ is at the center of the image
plane. Moreover, $o^{\textrm{i}}$ is termed as the principal point,
whose pixel coordinate is $\left(u_{0},v_{0}\right)^{\mathrm{T}}$.
In contrast, $o^{\textrm{c}}$ is termed as the camera optical center.
Furthermore, $o^{\textrm{i}}$ and $o^{\textrm{c}}$ are on the optical
axis. The distance between $o^{\textrm{c}}$ and $o^{\textrm{i}}$
is the focal length $f$, and thus the $z$-coordinate of the image
plane in CCS is $z^{\mathrm{c}}=f$.

A VLC-enabled rectangular LED luminaire is constructed by four vertices
$P_{i}$ ($i\in\left\{ 1,2,3,4\right\} $) mounted on the ceiling.
The four 3D reference lines $L_{ij}$ ($i,j\in\left\{ 1,2,3,4\right\} ,i\neq j$)
are the edges of the luminaire. In addition, $P_{i}^{\textrm{w}}=\left(x_{i}^{\textrm{w}},y_{i}^{\textrm{w}},z_{i}^{\textrm{w}}\right)^{\mathrm{T}}$
is the world coordinate of the $i$th vertex of the luminaire, which
is assumed to be known at the transmitter and can be transmitted by
VLC as the time-domain information \cite{yang2016enhanced,yang2019relay}.
Moreover, the unit normal vector of the luminaire in WCS, $\mathbf{n}_{\mathrm{LED}}^{\textrm{w}}$,
can be calculated by the world coordinates of the luminaire's vertices.

On the other hand, the receiver is a standard pinhole camera which
is not coplanar with the luminaire. Therefore, the transmitter and
the receiver produced a rectangular pyramid $o^{\textrm{c}}-P_{1}P_{2}P_{3}P_{4}$
which contains many space-domain information. In the rectangular pyramid
$o^{\textrm{c}}-P_{1}P_{2}P_{3}P_{4}$, the rectangle $P_{1}P_{2}P_{3}P_{4}$
is called the base face. Meanwhile, we define $\Pi_{ij}$ as the lateral
face determined by the vertices $P_{i}$, $P_{j}$ ($i,j\in\left\{ 1,2,3,4\right\} ,i\neq j$)
and $o^{\textrm{c}}$. In addition, $P_{i}$ is the $i$th vertex
and $o^{\textrm{c}}$ is called the apex. In the camera, $p_{i}$
is the projection of $P_{i}$ on the image plane. Moreover, $l_{ij}$
is the 2D projection on the image plane of $L_{ij}$. Note that many
existing PnL algorithms assume that the 3D-2D correspondences $\left(L_{ij}\iff l_{ij}\right)$
are known in advance, which is too ideal in practice \cite{ICRA2018Manhattan}.
In contrast, in this work, the 3D-2D correspondences are unknown.
To estimate the pose and position of the receiver without the 3D-2D
correspondences, the camera is used to simultaneously capture the
time- and space-domain information.

The pixel coordinate of $p_{i}$ is denoted by $p_{i}^{\textrm{p}}=\left(u_{i}^{\textrm{p}},v_{i}^{\textrm{p}}\right)^{\mathrm{T}}$,
and this coordinate can be obtained by the camera through image processing
\cite{li2018vlc}. Based on the single-view geometry theory, the image
coordinate of $p_{i}$, $p_{i}^{\textrm{i}}=\left(x_{i}^{\textrm{i}},y_{i}^{\textrm{i}}\right)^{\mathrm{T}}$,
can be obtained as follows:
\begin{equation}
p_{i}^{\textrm{i}}=\begin{bmatrix}d_{x}\\
d_{y}
\end{bmatrix}p_{i}^{\textrm{p}}-\begin{bmatrix}u_{0}d_{x}\\
v_{0}d_{y}
\end{bmatrix},\label{eq:1}
\end{equation}
where $d_{x}$ and $d_{y}$ are the physical size of each pixel in
the $x$ and $y$ directions on the image plane, respectively. The
camera's intrinsic parameters, including $\left(u_{0},v_{0}\right)^{\mathrm{T}}$
and the focal ratio $f_{u}=\frac{f}{d_{x}}$ and $f_{v}=\frac{f}{d_{y}}$,
can be calibrated in advance \cite{bai2019camera}. The transformation
from CCS to WCS can be expressed as follows \cite{xu2016pose}:
\begin{equation}
P^{\textrm{w}}=\mathbf{R}_{\mathrm{c}}^{\mathrm{w}}\cdot P^{\textrm{c}}+\mathbf{t_{\mathrm{c}}^{\mathrm{w}}},\label{eq:3}
\end{equation}
where $P^{\textrm{w}}$ and $P^{\textrm{c}}$ are the world and camera
coordinates of the same object, respectively. In addition, $\mathbf{R}_{\mathrm{c}}^{\mathrm{w}}$
and $\mathbf{t_{\mathrm{c}}^{\mathrm{w}}}\in\mathbb{R}^{3}$ denote
the pose and the position of the camera in WCS, respectively. The
task of the localization is to find out $\mathbf{R}_{\mathrm{c}}^{\mathrm{w}}$
and $\mathbf{t_{\mathrm{c}}^{\mathrm{w}}}$.

\vspace{-0.4cm}

\section{\label{sec:luminaire information}Calculating The Luminaire Information
In CCS}

In this section, the information of the luminaire in CCS, including
its normal vector $\mathbf{n}_{\mathrm{LED}}^{\mathrm{c}}$ and its
vertices' coordinates $P_{i}^{\textrm{c}}$, $i\in\left\{ 1,2,3,4\right\} $,
is calculated based on the space-domain information in two steps.
In the first step, $\mathbf{n}_{\mathrm{LED}}^{\mathrm{c}}$ is estimated
based on the plane and solid geometry theory. Then, based on $\mathbf{n}_{\mathrm{LED}}^{\mathrm{c}}$,
$P_{i}^{\textrm{c}}$, $i\in\left\{ 1,2,3,4\right\} $ are estimated
by the solid geometry theory.

\vspace{-0.5cm}

\subsection{\label{subsec:normal vector of LED in CCS}The Normal Vector Of The
Luminaire In CCS}

In ICS, the point-normal form equation of a given $l_{ij}$ can be
expressed as \cite{xiaojian2008analytic}:
\begin{equation}
x^{\textrm{i}}\cos\phi_{l_{ij}}+y^{\textrm{i}}\sin\phi_{l_{ij}}=\rho_{l_{ij}},\label{eq:50}
\end{equation}
where $\left(x^{\textrm{i}},y^{\textrm{i}}\right)^{\mathrm{T}}$ is
the image coordinate of a point on $l_{ij}$ which can be obtained
by the single-view geometry theory , $\phi_{l_{ij}}$ is the rotate
angle from $y^{\textrm{i}}$-axis to $l_{ij}$ in anticlockwise direction,
and $\rho_{l_{ij}}$ is the distance from $o^{\textrm{i}}$ to $l_{ij}$.
Since $p_{i}$ and $p_{j}$ are on $l_{ij}$, $\phi_{l_{ij}}$ and
$\rho_{l_{ij}}$ can be obtained based on the image coordinates of
$p_{i}$ and $p_{j}$. From (\ref{eq:3}), there are two points whose
image coordinates are $p_{l_{ij},1}^{\textrm{i}}=\left(\frac{\rho_{l_{ij}}}{\cos\phi_{l_{ij}}},0\right)^{\mathrm{T}}$
and $p_{l_{ij},2}^{\textrm{i}}=\left(0,\frac{\rho_{l_{ij}}}{\sin\phi_{l_{ij}}}\right)^{\mathrm{T}}$
are on $l_{ij}$. Since the two points are also on the image plane,
their camera coordinates are $p_{l_{ij},1}^{\textrm{c}}=\left(\frac{\rho_{l_{ij}}}{\cos\phi_{l_{ij}}},0,f\right)^{\mathrm{T}}$
and $p_{l_{ij},2}^{\textrm{c}}=\left(0,\frac{\rho_{l_{ij}}}{\sin\phi_{l_{ij}}},f\right)^{\mathrm{T}}$.
Since the two points and $o^{\textrm{c}}=\left(0,0,0\right)^{\mathrm{T}}$
are on $\Pi_{ij}$, we can represent $\Pi_{ij}$ in CCS in the general
form as:

\begin{equation}
A_{\Pi_{ij}}x^{\mathrm{c}}+B_{\Pi_{ij}}y^{\mathrm{c}}+C_{\Pi_{ij}}z^{\mathrm{c}}=0,\label{eq:8}
\end{equation}
where $A_{\Pi_{ij}}=f\cos\phi_{l_{ij}}$, $B_{\Pi_{ij}}=f\sin\phi_{l_{ij}}$
and $C_{\Pi_{ij}}=-\rho_{l_{ij}}$.

In CCS, the general form equation of the rectangle $P_{1}P_{2}P_{3}P_{4}$
can be expressed as:
\begin{equation}
A_{\mathrm{LED}}x^{\mathrm{c}}+B_{\mathrm{LED}}y^{\mathrm{c}}+C_{\mathrm{LED}}z^{\mathrm{c}}=1,\label{eq:12}
\end{equation}
where $A_{\mathrm{LED}}$, $B_{\mathrm{LED}}$ and $C_{\mathrm{LED}}$
are unknown constants. From (\ref{eq:12}), the normal vector of the
rectangle $P_{1}P_{2}P_{3}P_{4}$ can be expressed by $\left(A_{\mathrm{LED}},B_{\mathrm{LED}},C_{\mathrm{LED}}\right)^{\mathrm{T}}$.
In CCS, let $\mathbf{n}_{\Pi_{ij}}^{\mathrm{c}}=\left(A_{\Pi_{ij}},B_{\Pi_{ij}},C_{\Pi_{ij}}\right)^{\mathrm{T}}$
($i,j\in\left\{ 1,2,3,4\right\} ,i\neq j$) denotes the normal vector
of $\Pi_{ij}$ and $\mathbf{v}_{L_{ij}}^{\mathrm{c}}\in\mathbb{R}^{3}$
($i,j\in\left\{ 1,2,3,4\right\} ,i\neq j$) denotes the direction
vector of $L_{ij}$. Since $L_{ij}$ is the intersection line of the
rectangle $P_{1}P_{2}P_{3}P_{4}$ and $\Pi_{ij}$, $\mathbf{v}_{L_{ij}}^{\mathrm{c}}$
can be calculated as $\mathbf{v}_{L_{ij}}^{\mathrm{c}}=\left(A_{\mathrm{LED}},B_{\mathrm{LED}},C_{\mathrm{LED}}\right)^{\mathrm{T}}\times\mathbf{n}_{\Pi_{ij}}^{\mathrm{c}}$.
Based on the solid geometry, we have:
\begin{equation}
\begin{cases}
\mathbf{v}_{L_{34}}^{\mathrm{c}}\cdot\mathbf{n}_{\Pi_{12}}^{\mathrm{c}}=0\\
\mathbf{v}_{L_{41}}^{\mathrm{c}}\cdot\mathbf{n}_{\Pi_{23}}^{\mathrm{c}}=0.
\end{cases}\label{eq:11}
\end{equation}
Define $m=\frac{A_{\mathrm{LED}}}{C_{\mathrm{LED}}}$ and $n=\frac{B_{\mathrm{LED}}}{C_{\mathrm{LED}}}$,
and we can obtain $m$ and $n$ as the functions of $A_{\Pi_{ij}}$,
$B_{\Pi_{ij}}$ and $C_{\Pi_{ij}}$ by solving (\ref{eq:11}). Therefore,
the normalized normal vector of the rectangle $P_{1}P_{2}P_{3}P_{4}$
(i.e., the orientation of the luminaire) in CCS can be expressed as:
\begin{equation}
\mathbf{n}_{\mathrm{LED}}^{\mathrm{c}}=\left(\cos\alpha,\cos\beta,\cos\gamma\right)^{\mathrm{T}},\label{eq:13}
\end{equation}
where:
\begin{equation}
\begin{cases}
\cos\alpha=\frac{m}{\sqrt{m^{2}+n^{2}+1}}\\
\cos\beta=\frac{n}{\sqrt{m^{2}+n^{2}+1}}\\
\cos\gamma=\frac{1}{\sqrt{m^{2}+n^{2}+1}}.
\end{cases}\label{eq:14}
\end{equation}
\vspace{-0.6cm}

\subsection{\label{subsec:led camera coordinate}Camera Coordinates Of The Luminaire's
Vertices}

Since $P_{1}$ is the intersection point of the rectangle $P_{1}P_{2}P_{3}P_{4}$,
$\Pi_{12}$ and $\Pi_{41}$, its camera coordinate can be calculated
as $P_{1}^{\textrm{c}}=\frac{\mathbf{M}_{P_{1}}}{C_{\mathrm{LED}}}$,
where:
\begin{equation}
\mathbf{M}_{P_{1}}=\begin{bmatrix}m & n & 1\\
A_{\Pi_{12}} & B_{\Pi_{12}} & C_{\Pi_{12}}\\
A_{\Pi_{41}} & B_{\Pi_{41}} & C_{\Pi_{41}}
\end{bmatrix}^{-1}\cdot\begin{bmatrix}1\\
0\\
0
\end{bmatrix}.\label{eq:100}
\end{equation}
The other three $\mathbf{M}_{P_{i}}$ ($i\in\left\{ 2,3,4\right\} $)
can be calculated in the similar method of (\ref{eq:100}). In general,
the camera coordinate $P_{i}^{\textrm{c}}$ ($i\in\left\{ 1,2,3,4\right\} $)
can be calculated as follows:
\begin{equation}
P_{i}^{\textrm{c}}=\frac{\mathbf{M}_{P_{i}}}{C_{\mathrm{LED}}}.\label{eq:37}
\end{equation}
From (\ref{eq:37}), we can observe that $P_{i}^{\textrm{c}}$ can
be represented according to $C_{\mathrm{LED}}$. Next, we will calculate
$C_{\mathrm{LED}}$ based on the solid geometry.

The volume of the rectangular pyramid $o^{\textrm{c}}-P_{1}P_{2}P_{3}P_{4}$
can be calculates as $V=\frac{1}{3}SH$, where $S$ is the area of
the luminaire and is known in advance. Additionally, $H=\frac{1}{C_{\mathrm{LED}}\sqrt{m^{2}+n^{2}+1}}$
is the distance from $o^{\textrm{c}}$ to the rectangle $P_{1}P_{2}P_{3}P_{4}$.
For the triangular pyramid $o^{\textrm{c}}-P_{1}P_{2}P_{3}$, its
volume can be calculated as follows:
\begin{equation}
V_{1}=\frac{1}{6}\left|\det(\mathbf{M}_{V_{1}})\right|,\label{eq:19}
\end{equation}
where $\mathbf{M}_{V_{1}}=\left[P_{1}^{\textrm{c}},P_{2}^{\textrm{c}},P_{3}^{\textrm{c}}\right]^{\mathrm{T}}$.
Substituting (\ref{eq:37}) into (\ref{eq:19}), we have $V_{1}=\frac{q_{1}}{C_{\mathrm{LED}}^{3}}$,
where $q_{1}=\frac{1}{6}\left|\det(\mathbf{M}_{q_{1}})\right|$, where
$\mathbf{M}_{q_{1}}=\left[\mathbf{M}_{P_{1}},\mathbf{M}_{P_{2}},\mathbf{M}_{P_{3}}\right]^{\mathrm{T}}$.
The volumes of the other three triangular pyramid $o^{\textrm{c}}-P_{2}P_{3}P_{4}$,
$o^{\textrm{c}}-P_{3}P_{4}P_{1}$ and $o^{\textrm{c}}-P_{4}P_{1}P_{2}$,
denoted by $V_{2}$, $V_{3}$ and $V_{4}$, respectively, can be obtained
in the same way. Since $V=\frac{1}{2}\sum_{i=1}^{4}V_{i}$, $C_{\mathrm{LED}}$
can be calculated as follows:
\begin{equation}
C_{\mathrm{LED}}=\sqrt{\frac{3\sum_{i=1}^{4}q_{i}\cdot\sqrt{m^{2}+n^{2}+1}}{2S}}.\label{eq:21}
\end{equation}
Substituting (\ref{eq:21}) into (\ref{eq:37}), $P_{i}^{\textrm{c}}$
($i\in\left\{ 1,2,3,4\right\} $) can be obtained.

\vspace{-0.3cm}

\section{\label{sec:V-P4L-SH}The Basic Algorithm of V-P4L}

In this section, the basic algorithm of V-P4L is proposed for scenarios
where LEDs have the same height. The basic algorithm of V-P4L contains
three steps. In the first step, based on the orientation information
of the luminaire estimated in Section \ref{sec:luminaire information},
the rotation angles corresponding to the $x^{\mathrm{c}}-$axis and
$y^{\mathrm{c}}-$axis can be obtained by the single-view geometry
theory. Then, based on the LLS method and the single-view geometry
theory, the basic algorithm of V-P4L can properly match the 3D-2D
correspondences, and obtain the rotation angles corresponding to the
$z^{\mathrm{c}}-$axis and the 2D coordinate of the camera. Finally,
based on the single-view geometry theory, V-P4L can estimate the $z$-coordinate
of the camera.

\subsection{\label{subsec:xy_angles}Calculate the rotation angles corresponding
to the $x^{\mathrm{c}}-$axis and $y^{\mathrm{c}}-$axis}

Let $\mathbf{R}_{X}$, $\mathbf{R}_{Y}$ and $\mathbf{R}_{Z}$ denote
the rotation matrices of WCS along the $x^{\mathrm{c}}-$axis, $y^{\mathrm{c}}-$axis
and $z^{\mathrm{c}}-$axis, respectively. Given $\mathbf{R}_{X}$,
$\mathbf{R}_{Y}$ and $\mathbf{R}_{Z}$ as follows \cite{Taylor1986rotation}:
\begin{equation}
\mathbf{R}_{X}=\begin{bmatrix}1 & 0 & 0\\
0 & \cos\varphi & -\sin\varphi\\
0 & \sin\varphi & \cos\varphi
\end{bmatrix},\label{eq:22}
\end{equation}
\begin{equation}
\mathbf{R}_{Y}=\begin{bmatrix}\cos\theta & 0 & \sin\theta\\
0 & 1 & 0\\
-\sin\theta & 0 & \cos\theta
\end{bmatrix},\label{eq:23}
\end{equation}
and
\begin{equation}
\mathbf{R}_{Z}=\begin{bmatrix}\cos\psi & -\sin\psi & 0\\
\sin\psi & \cos\psi & 0\\
0 & 0 & 1
\end{bmatrix},\label{eq:24}
\end{equation}
where $\varphi\in\left(-\frac{\pi}{2},\frac{\pi}{2}\right]$, $\theta\in\left(-\frac{\pi}{2},\frac{\pi}{2}\right]$
and $\psi\in\left(-\pi,\pi\right]$ are the unknown Euler angles corresponding
to the $x^{\mathrm{c}}-$axis, $y^{\mathrm{c}}-$axis and $z^{\mathrm{c}}-$axis,
respectively, the rotation matrix $\mathbf{R}_{\mathrm{c}}^{\mathrm{w}}$
from CCS to WCS can be given as \cite{Taylor1986rotation}:
\begin{equation}
\mathbf{R}_{\mathrm{c}}^{\mathrm{w}}=\mathbf{R}_{Z}\mathbf{R}_{Y}\mathbf{R}_{X}.\label{eq:25}
\end{equation}
In this section, the basic algorithm of V-P4L is proposed for scenarios
where LEDs have the same height. Therefore, the normal vector of the
luminaire can be denoted by $\mathbf{n}_{\unit{LED}}^{\textrm{w}}=\left(0,0,1\right)^{\mathrm{T}}$.
Based on the single-view geometry theory, the relationship between
$\mathbf{n}_{\unit{LED}}^{\textrm{w}}=\left(0,0,1\right)^{\mathrm{T}}$
and $\mathbf{n}_{\mathrm{LED}}^{\mathrm{c}}=\left(\cos\alpha,\cos\beta,\cos\gamma\right)^{\mathrm{T}}$
can be given as \cite{xu2016pose}:
\begin{equation}
\mathbf{n}_{\mathrm{LED}}^{\mathrm{w}}=\mathbf{R}_{\mathrm{c}}^{\mathrm{w}}\cdot\mathbf{n}_{\mathrm{LED}}^{\mathrm{c}}.\label{eq:40}
\end{equation}
Therefore, we have:
\begin{equation}
\begin{cases}
\cos\alpha=-\sin\theta\\
\cos\beta=\cos\theta\cdot\sin\varphi\\
\cos\gamma=\cos\theta\cdot\cos\varphi.
\end{cases}\label{eq:29}
\end{equation}
The estimated rotation angles $\hat{\varphi}$ and $\hat{\theta}$
can be obtained by solving (\ref{eq:29}).

\subsection{\label{subsec:z_angle tx ty}Calculate the rotation angles corresponding
to the $z^{\mathrm{c}}-$axis and the 2D coordinate of the camera}

Based on the single-view geometry theory, the relationship between
$P_{i}^{\textrm{c}}$ ($i\in\left\{ 1,2,3,4\right\} $) and $P_{i}^{\textrm{w}}=\left(x_{i}^{\textrm{w}},y_{i}^{\textrm{w}},z_{i}^{\textrm{w}}\right)^{\mathrm{T}}$
can be given as \cite{xu2016pose}:
\begin{equation}
P_{i}^{\textrm{w}}=\mathbf{R}_{\mathrm{c}}^{\mathrm{w}}\cdot P_{i}^{\textrm{c}}+\mathbf{t_{\mathrm{c}}^{\mathrm{w}}},\label{eq:26}
\end{equation}
where $P_{i}^{\textrm{w}}$ is known in advance and can be obtained
by the camera as the time-domain information. In addition, $P_{i}^{\textrm{c}}$
is the space-domain information that is estimated in Subsection \ref{subsec:led camera coordinate}.
Moreover, $\mathbf{t_{\mathrm{c}}^{\mathrm{w}}}=\left(t_{x},t_{y},t_{z}\right)^{\mathrm{T}}$
is the 3D world coordinate of the camera. In (\ref{eq:26}), there
are four unknown parameters $\psi$, $t_{x}$, $t_{y}$ and $t_{z}$.
If the 3D-2D correspondences are known in advance, we can easily obtain
the four unknown parameters with the four vertices' world and camera
coordinates. However, as analyzed in Section \ref{sec:intro}, the
3D-2D correspondences are unknown for practical considerations. In
this paper, we can calculate these parameters based on the LEDs' information
in both time and space domains. For mathematical analysis, we define:
\begin{equation}
\mathbf{R}_{Y}\mathbf{R}_{X}=\left[\begin{array}{ccc}
\cos\theta & \sin\theta\cdot\sin\varphi & \sin\theta\cdot\cos\varphi\\
0 & \cos\varphi & -\sin\varphi\\
-\sin\theta & \cos\theta\cdot\sin\varphi & \cos\theta\cdot\cos\varphi
\end{array}\right]=\begin{bmatrix}a_{1} & a_{2} & a_{3}\\
b_{1} & b_{2} & b_{3}\\
c_{1} & c_{2} & c_{3}
\end{bmatrix},\label{eq:35}
\end{equation}
and thus we can rewrite (\ref{eq:26}) as follows:
\begin{equation}
P_{i}^{\textrm{w}}=\begin{bmatrix}\cos\psi & -\sin\psi & 0\\
\sin\psi & \cos\psi & 0\\
0 & 0 & 1
\end{bmatrix}\cdot\begin{bmatrix}a_{1} & a_{2} & a_{3}\\
b_{1} & b_{2} & b_{3}\\
c_{1} & c_{2} & c_{3}
\end{bmatrix}\cdot P_{i}^{\textrm{c}}+\mathbf{t_{\mathrm{c}}^{\mathrm{w}}}.\label{eq:27}
\end{equation}
The three unknown parameters $\psi$, $t_{x}$ and $t_{y}$ in (\ref{eq:27})
can be calculated by the LLS estimator, which can be expressed in
a matrix form as follows:
\begin{equation}
\mathbf{A_{\mathit{rs}}}\cdot\mathbf{x}=\mathbf{b_{\mathit{ij}}},\label{eq:31}
\end{equation}
where $\mathbf{A}_{\mathit{rs}}=\left[\mathbf{A}_{r};\mathbf{A}_{s}\right]$
and $\mathbf{b_{\mathit{ij}}}=\left[\mathbf{b}_{i};\mathbf{b}_{j}\right]$
($r,s,i,j\in\left\{ 1,2,3,4\right\} ,i\neq j,r\neq s$), where:
\begin{equation}
\mathbf{A}_{r}=\begin{bmatrix}\left[a_{1},a_{2},a_{3}\right]^{\mathrm{T}}\cdot P_{r}^{\textrm{c}} & -\left[b_{1},b_{2},b_{3}\right]^{\mathrm{T}}\cdot P_{r}^{\textrm{c}} & 1 & 0\\
\left[b_{1},b_{2},b_{3}\right]^{\mathrm{T}}\cdot P_{r}^{\textrm{c}} & \left[a_{1},a_{2},a_{3}\right]^{\mathrm{T}}\cdot P_{r}^{\textrm{c}} & 0 & 1
\end{bmatrix}(r\in\left\{ 1,2,3,4\right\} ),\label{eq:31-1}
\end{equation}

\begin{equation}
\mathbf{x}=\left[\cos\psi,\sin\psi,t_{x},t_{y}\right]^{\mathrm{T}},\label{eq:31-2}
\end{equation}
and
\begin{equation}
\mathbf{b}_{i}=\left[x_{i}^{\textrm{w}},y_{i}^{\textrm{w}}\right]^{\mathrm{T}}(i\in\left\{ 1,2,3,4\right\} ).\label{eq:31-3}
\end{equation}
Therefore, the unknown parameters can be given by:
\begin{equation}
\mathbf{\hat{x}=(A_{\mathit{rs}}^{\mathrm{T}}A_{\mathit{rs}})^{\mathrm{-1}}A_{\mathit{rs}}^{\mathrm{T}}b_{\mathit{ij}}},\label{eq:32}
\end{equation}
where $\hat{\mathbf{x}}=\left[\hat{\cos\psi},\hat{\sin\psi},\hat{t}_{x},\hat{t}_{y}\right]^{\mathrm{T}}$
is the estimate of $\mathbf{x}$.

Since the 3D-2D correspondences are not known in advance, given a
certain $\mathbf{A}_{\mathit{rs}}$ and $\mathbf{b_{\mathit{ij}}}$
($r,s,i,j\in\left\{ 1,2,3,4\right\} ,r\neq s,i\neq j$), we cannot
obtain their exact correspondence relationship. Fortunately, there
are four $\mathbf{A}_{r}$ ($r\in\left\{ 1,2,3,4\right\} $) and $\mathbf{b}_{i}$
($i\in\left\{ 1,2,3,4\right\} $), and that means there are only $\mathrm{C}_{4}^{2}=6$
different $\mathbf{A}_{\mathit{rs}}$ and $\mathbf{b_{\mathit{ij}}}$
($r,s,i,j\in\left\{ 1,2,3,4\right\} ,r\neq s,i\neq j$). Therefore,
for each $\mathbf{b_{\mathit{ij}}}=\left[\mathbf{b}_{i},\mathbf{b}_{j}\right]^{\mathrm{T}}$
($i,j\in\left\{ 1,2,3,4\right\} $, $i\neq j$), we can obtain 6 candidate
solutions corresponding to 6 $\mathbf{A}_{\mathit{rs}}$ ($r,s\in\left\{ 1,2,3,4\right\} ,r\neq s$),
and one of which is $\hat{\mathbf{x}}_{ij}$, where $\hat{\mathbf{x}}_{ij}$
represents the exact $\hat{\mathbf{x}}$ corresponding to $\mathbf{b_{\mathit{ij}}}$.
Therefore, we can obtain total 36 solutions which can further be separated
into 6 groups according to 6 different $\mathbf{b_{\mathit{ij}}}$.
To obtain a reasonable solution, here we propose a strategy that estimates
$\hat{\mathbf{x}}$ by averaging the 6 closest solutions (i.e., 6
$\hat{\mathbf{x}}_{ij}$) in the 6 groups of solutions, which can
be expressed as follows:
\begin{equation}
\hat{\mathbf{x}}=\frac{1}{6}\sum_{i=1}^{4}\sum_{j=1,j>i}^{4}\hat{\mathbf{x}}_{ij}.\label{eq:39}
\end{equation}
This strategy will be verified in simulations. In this way, based
on the information in both time and space domains, the basic algorithm
of V-P4L can properly match the 3D-2D correspondences, and obtain
the rotation angles corresponding to the $z^{\mathrm{c}}-$axis $\psi$
and the 2D coordinate of the camera $\left(t_{x},t_{y}\right)^{\mathrm{T}}$.

\subsection{\label{subsec:tz}Calculate the $z$-coordinate of the camera}

In Subsection \ref{subsec:z_angle tx ty}, we have obtained $\psi$
and $\left(t_{x},t_{y}\right)^{\mathrm{T}}$. In (\ref{eq:27}), there
is still one unknown parameter $t_{z}$. In this section, the basic
algorithm of V-P4L is proposed for scenarios where LEDs have the same
height, i.e., $z_{1}^{\unit{w}}=z_{2}^{\unit{w}}=z_{3}^{\unit{w}}=z_{4}^{\unit{w}}$.
Based on the single-view geometry theory, we can obtain the relationship
between $z_{i}^{\textrm{w}}$ and $t_{z}$ from (\ref{eq:27}) as:
\begin{equation}
z_{i}^{\textrm{w}}=\left[c_{1},c_{2},c_{3}\right]^{\mathrm{T}}\cdot P_{i}^{\textrm{c}}+t_{z}.\label{eq:33}
\end{equation}
The estimated $z$-coordinate of the camera in WCS $\hat{t}_{z}$
can be calculated as follows:
\begin{equation}
\hat{t}_{z}=\frac{1}{4}\left(\sum_{i=1}^{4}z_{i}^{\textrm{w}}-\sum_{i=1}^{4}\left[\hat{c}_{1},\hat{c}_{2},\hat{c}_{3}\right]^{\mathrm{T}}\cdot P_{i}^{\textrm{c}}\right),\label{eq:30}
\end{equation}
where $\hat{c}_{k}$ $\left(k\in\left\{ 1,2,3\right\} \right)$ is
the estimate of $c_{k}$.

In this way, the estimated pose $\mathbf{\hat{R}}_{\mathrm{c}}^{\mathrm{w}}$
and position of the camera $\mathbf{\hat{t}_{\mathrm{c}}^{\mathrm{w}}}$
can be obtained without the ideal 2D-3D correspondence assumption.

\vspace{-0cm}

\section{\label{sec:V-P4L-DH}The Correction algorithm of V-P4L}

Most of existing studies including the basic algorithm of V-P4L proposed
in Subsection \ref{sec:V-P4L-SH} assume that LEDs have the same height
\cite{bai2020enhanced,li2014epsilon}. However, this may not always
be true in practice. For instance, ceilings may be tilted due to the
imperfect decoration or deliberate design. In these scenarios, the
localization accuracy can be significantly degraded. Therefore, in
this subsection, based on the basic algorithm of V-P4L, we propose
a correction algorithm of V-P4L for the scenarios where LEDs have
different heights, V-P4L-DH. Based on the single-view geometry theory
and the LLS method, V-P4L-DH can properly match the 3D-2D correspondences
and obtain the 2D position of the camera. Based on the 2D localization,
V-P4L-DH can achieve 3D localization by a simple optimization method.

\subsection{\label{subsec:DH-2D}2D Localization}

For 2D-localization case where the $z$-coordinate of the camera $t_{z}$
is known in advance, based on the single-view geometry theory, (\ref{eq:33})
can be rewritten as follows:
\begin{equation}
\left[c_{1},c_{2},c_{3}\right]^{\mathrm{T}}\cdot P_{i}^{\textrm{c}}=z_{j}^{\textrm{w}}-t_{z},\label{eq:42}
\end{equation}
where $i,j\in\left\{ 1,2,3,4\right\} $. Since the 3D-2D correspondence
are not known in advance, given a certain $P_{i}^{\textrm{c}}$ and
$z_{j}^{\textrm{w}}$, we do not know their correspondence relationship.
Fortunately, there are four $P_{i}^{\textrm{c}}$ and $z_{j}^{\textrm{w}}$,
and thus we can estimate $c_{k}$ $\left(k\in\left\{ 1,2,3\right\} \right)$
using the same LLS method of solving (\ref{eq:32}). Then, from (\ref{eq:35}),
we have:
\begin{equation}
\begin{cases}
\hat{c}_{1}=-\sin\theta\\
\hat{c}_{2}=\cos\theta\cdot\sin\varphi\\
\hat{c}_{3}=\cos\theta\cdot\cos\varphi,
\end{cases}\label{eq:43}
\end{equation}
where $\hat{c}_{k}$ $\left(k\in\left\{ 1,2,3\right\} \right)$ is
the estimate of $c_{k}$. The estimated rotation angles corresponding
to the $x^{\mathrm{c}}-$axis $\hat{\varphi}$ and $y^{\mathrm{c}}-$axis
$\hat{\theta}$ can be obtained by solving (\ref{eq:43}). Then, the
pose of the camera $\mathbf{\hat{R}}_{\mathrm{c}}^{\mathrm{w}}$ and
2D position of the camera $\mathbf{\hat{t}_{\mathrm{c,2D}}^{\mathrm{w}}}=\left(\hat{t}_{x},\hat{t}_{y}\right)^{\mathrm{T}}$
can be obtained according to (\ref{eq:35})$-$(\ref{eq:39}), where
$\hat{t}_{x}$ and $\hat{t}_{y}$ are the estimated $x$ and $y$-coordinates
of the camera, respectively.

\subsection{\label{subsec:DH-3D}3D Localization}

For 3D-localization case, the $z$-coordinate of the camera $t_{z}$
is not known in advance. For indoor scenario, the range of $t_{z}$
must be $[0,H_{\mathrm{m}})$, where $H_{\mathrm{m}}$ is the maximum
height of the room. Based on the above 2D-localizatoin algorithm,
for different $t_{z}\in[0,H_{\mathrm{m}})$, we can obtain different
$\mathbf{\hat{R}}_{\mathrm{c}}^{\mathrm{w}}\left(t_{z}\right)$ and
$\mathbf{\hat{t}_{\mathrm{c,2D}}^{\mathrm{w}}}\left(t_{z}\right)$.
Based on the estimated normal vector of the luminaire in CCS $\mathbf{n}_{\mathrm{LED}}^{\mathrm{c}}$,
we have \cite{xu2016pose}:
\begin{equation}
\mathbf{\hat{n}}_{\mathrm{LED}}^{\mathrm{w}}\left(t_{z}\right)=\mathbf{\hat{R}}_{\mathrm{c}}^{\mathrm{w}}\left(t_{z}\right)\cdot\mathbf{n}_{\mathrm{LED}}^{\mathrm{c}},\label{eq:41}
\end{equation}
where $\mathbf{\hat{n}}_{\mathrm{LED}}^{\mathrm{w}}\left(t_{z}\right)$
denotes the estimated normal vector of the luminaire in WCS when the
$z$-coordinate of the camera is $t_{z}$. Since the world coordinates
of the luminaire's vertices $\mathbf{s}_{i}^{\textrm{w}}$ ($i\in\left\{ 1,2,3,4\right\} $)
are known in advance, the actual normal vector of the luminaire in
WCS $\mathbf{n}_{\unit{LED}}^{\textrm{w}}$ can be calculated as follows:
\begin{equation}
\mathbf{n}_{\unit{LED}}^{\textrm{w}}=\left(P_{i}^{\textrm{w}}-P_{j}^{\textrm{w}}\right)\times\left(P_{i}^{\textrm{w}}-P_{k}^{\textrm{w}}\right),\label{eq:44}
\end{equation}
where $i,j,k\in\left\{ 1,2,3,4\right\} ,i\neq j\neq k$. Therefore,
the difference between the estimated and actual normal vectors of
the luminaire in WCS can be given as:
\begin{equation}
\Delta G\left(t_{z}\right)=\left\Vert \mathbf{n}_{\unit{LED}}^{\textrm{w}}-\mathbf{\hat{n}}_{\unit{LED}}^{\textrm{w}}\left(t_{z}\right)\right\Vert _{2}.\label{eq:45}
\end{equation}
The estimated $z$-coordinate of the camera $\hat{t}_{z}$ can be
obtained by the minimum $\Delta G\left(t_{z}\right)$, i.e.:
\begin{equation}
\hat{t}_{z}=\arg\,\min_{t_{z}}\Delta G\left(t_{z}\right).\label{eq:46}
\end{equation}

To reduce the complexity of V-P4L-DH, we propose a $n$-step segmentation
optimization strategy for V-P4L-DH. In the first step, we divide the
range of $t_{z}$ into $N$ segments evenly, and set:
\begin{equation}
t_{z}\in\left\{ 0,\frac{H_{\mathrm{m}}}{N},2\frac{H_{\mathrm{m}}}{N},3\frac{H_{\mathrm{m}}}{N},\ldots,H_{\mathrm{m}}\right\} ,\label{eq:61}
\end{equation}
i.e., we set the interval between adjacent $t_{z}$ as $\varepsilon_{1}\triangleq\frac{H_{\mathrm{m}}}{N}$.
Substituting all the $t_{z}$ into (\ref{eq:45}), we have $\Delta G\left(t_{z}\right)=\left\{ \Delta G\left(t_{z,0}\right),\Delta G\left(t_{z,1}\right),\ldots,\Delta G\left(t_{z,N}\right)\right\} $.
According to (\ref{eq:46}), we can find the minimal $\Delta G\left(t_{z}\right)$.
We denote the minimal $\Delta G\left(t_{z}\right)$ by $\Delta G\left(t_{z,i}\right)$
where $i\in\left\{ 0,1,\ldots,N\right\} $ is the index of the $t_{z}$
that corresponding to $\Delta G\left(t_{z,i}\right)$. In the second
step, we reduce the range of $t_{z}$ to $\left(\frac{H_{\mathrm{m}}}{N}\left(i-1\right),\frac{H_{\mathrm{m}}}{N}\left(i+1\right)\right)$.
We set:
\begin{equation}
t_{z}\in\left\{ \frac{H_{\mathrm{m}}}{N}\left(i-1\right)+\varepsilon_{2},\frac{H_{\mathrm{m}}}{N}\left(i-1\right)+2\varepsilon_{2},\ldots,\frac{H_{\mathrm{m}}}{N}\left(i+1\right)-\varepsilon_{2}\right\} ,\label{eq:60}
\end{equation}
i.e., the interval between adjacent $t_{z}$ is $\varepsilon_{2}$,
where $\varepsilon_{2}<\varepsilon_{1}$. We repeat the process of
the first step for the second step to the $n$th step until we can
obtain the precise $\hat{t}_{z}$ according to (\ref{eq:46}), and
obtain the optimal $\mathbf{\hat{R}}_{\mathrm{c}}^{\mathrm{w}}$ and
$\mathbf{\hat{t}_{\mathrm{c,2D}}^{\mathrm{w}}}$. Therefore, based
on the simple optimization method, we can obtain the pose and 3D position
of the receiver when LEDs have different heights.

\begin{figure}
\begin{centering}
\includegraphics[scale=0.6]{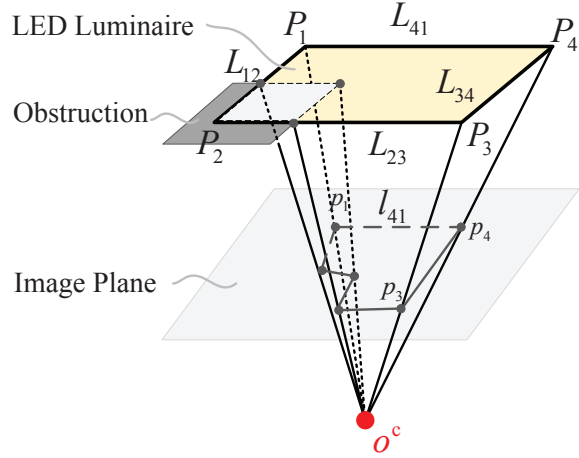}
\par\end{centering}
\caption{\label{fig:occlusion}A occlusion scenario where the projection of
$P_{2}$, $p_{2}$, is blocked by barriers and not on the image plane.}
\end{figure}

In this way, when LEDs have the same height, the basic algorithm of
V-P4L can be implemented for high accuracy at low complexity. When
LEDs have different heights, V-P4L-DH can be implemented for high
accuracy. In summary, V-P4L algorithm is elaborated in Algorithm 1.
Although V-P4L requires the LED luminaire to be a rectangle, it is
robust to partial occlusion, which is meaningful due to the limitation
on the camera's field of view. For instance, if the projection of
$P_{2}$, $p_{2}$, is blocked by barriers and not on the image plane
as shown in Fig. \ref{fig:occlusion}, the pixel coordinate of $p_{2}$
can be determined by the intersection of $l_{12}$ and $l_{23}$,
and thus V-P4L can be still successively implemented.

\begin{algorithm*}
\caption{V-P4L Algorithm.}

\textbf{Input:}

\:\qquad{}$P_{i}^{\textrm{w}}=\left(x_{i}^{\textrm{w}},y_{i}^{\textrm{w}},z_{i}^{\textrm{w}}\right)^{\mathrm{T}}$
($i\in\left\{ 1,2,3,4\right\} $);

\:\qquad{}$p_{j}^{\textrm{p}}=\left(u_{j}^{\textrm{p}},v_{j}^{\textrm{p}}\right)^{\mathrm{T}}$
($j\in\left\{ 1,2,3,4\right\} $);

\:\qquad{}$u_{0}$, $v_{0}$, $f$, $f_{u}$, $f_{v}$ and $\varepsilon_{1}$.

\begin{algorithmic}[1]

\STATE Calculate $p_{i}^{\textrm{i}}=\left(x_{i}^{\textrm{i}},y_{i}^{\textrm{i}}\right)^{\mathrm{T}}$($i\in\left\{ 1,2,3,4\right\} $)
according to (\ref{eq:1}).

\FOR {$i$ = 1 $\to$ $4$, $j$ = 1 $\to$ $4$ and $j\neq i$}

\STATE Calculate $\phi_{l_{ij}}$ and $\rho_{l_{ij}}$ by $p_{i}^{\textrm{i}}$
and $p_{j}^{\textrm{i}}$.

\STATE $A_{\Pi_{ij}}\leftarrow f\cos\phi_{l_{ij}}$, $B_{\Pi_{ij}}\leftarrow f\sin\phi_{l_{ij}}$
and $C_{\Pi_{ij}}\leftarrow-\rho_{l_{ij}}$, and then $\mathbf{n}_{\Pi_{ij}}^{\mathrm{c}}\leftarrow\left(A_{\Pi_{ij}},B_{\Pi_{ij}},C_{\Pi_{ij}}\right)^{\mathrm{T}}$.

\STATE $\mathbf{v}_{L_{ij}}^{\mathrm{c}}\leftarrow\left(A_{\mathrm{LED}},B_{\mathrm{LED}},C_{\mathrm{LED}}\right)^{\mathrm{T}}\times\mathbf{n}_{\Pi_{ij}}^{\mathrm{c}}$,
where $\left(A_{\mathrm{LED}},B_{\mathrm{LED}},C_{\mathrm{LED}}\right)^{\mathrm{T}}$
denotes the normal vector of the luminaire in CCS.

\ENDFOR

\STATE Define $m=\frac{A_{\mathrm{LED}}}{C_{\mathrm{LED}}}$ and
$n=\frac{B_{\mathrm{LED}}}{C_{\mathrm{LED}}}$. Calculate $m$ and
$n$ according to $\mathbf{v}_{L_{34}}^{\mathrm{c}}\cdot\mathbf{n}_{\Pi_{12}}^{\mathrm{c}}=0$
and $\mathbf{v}_{L_{41}}^{\mathrm{c}}\cdot\mathbf{n}_{\Pi_{23}}^{\mathrm{c}}=0$.

\STATE $\cos\alpha\leftarrow\frac{m}{\sqrt{m^{2}+n^{2}+1}}$, $\cos\beta\leftarrow\frac{n}{\sqrt{m^{2}+n^{2}+1}}$
and $\cos\gamma\leftarrow\frac{1}{\sqrt{m^{2}+n^{2}+1}}$, and then
$\mathbf{n}_{\mathrm{LED}}^{\mathrm{c}}\leftarrow\left(\cos\alpha,\cos\beta,\cos\gamma\right)^{\mathrm{T}}$.

\STATE Calculate $\mathbf{M}_{P_{1}}$ according to (\ref{eq:100}),
and $\mathbf{M}_{P_{i}}$ ($i\in\left\{ 2,3,4\right\} $) can be calculated
in the same way as $\mathbf{M}_{P_{1}}$.

\STATE $q_{1}\leftarrow\frac{1}{6}\left|\det(\mathbf{M}_{q_{1}})\right|$,
where $\mathbf{M}_{q_{1}}=\left[\mathbf{M}_{P_{1}},\mathbf{M}_{P_{2}},\mathbf{M}_{P_{3}}\right]^{\mathrm{T}}$,
and $q_{i}$ ($i\in\left\{ 2,3,4\right\} $) can be calculated in
the same way as $q_{1}$.

\STATE $C_{\mathrm{LED}}\leftarrow\sqrt{\frac{3\sum_{i=1}^{4}q_{i}\cdot\sqrt{m^{2}+n^{2}+1}}{2S}}$.

\STATE $P_{i}^{\textrm{c}}\leftarrow\frac{\mathbf{M}_{P_{i}}}{C_{\mathrm{LED}}}$
$\left(i\in\left\{ 1,2,3,4\right\} \right)$.

\IF {$z_{1}^{\textrm{w}}=z_{2}^{\textrm{w}}=z_{3}^{\textrm{w}}=z_{4}^{\textrm{w}}$}

\STATE $\mathbf{n}_{\unit{LED}}^{\textrm{w}}\leftarrow\left(0,0,1\right)^{\mathrm{T}}$.

\STATE Calculate $\hat{\varphi}$ and $\hat{\theta}$ according to
(\ref{eq:29}).

\STATE Calculate $\hat{\psi}$, $\hat{t}_{x}$ and $\hat{t}_{y}$
according to (\ref{eq:31}) - (\ref{eq:32}).

\STATE Calculate $\hat{t}_{z}$ according to (\ref{eq:30}) if $t_{z}$
is not known in advance.

\ELSE

\IF {$t_{z}$ is known in advance}

\STATE Calculate $\hat{c}_{i}$ $\left(i\in\left\{ 1,2,3\right\} \right)$
following the same method of solving (\ref{eq:32}).

\STATE Calculate $\hat{\varphi}$ and $\hat{\theta}$ according to
(\ref{eq:43}).

\STATE Calculate $\hat{\psi}$, $\hat{t}_{x}$ and $\hat{t}_{y}$
according to (\ref{eq:35}) - (\ref{eq:39}). Therefore, $\mathbf{\hat{R}}_{\mathrm{c}}^{\mathrm{w}}$
and $\mathbf{\hat{t}_{\mathrm{c,2D}}^{\mathrm{w}}}=\left(\hat{t}_{x},\hat{t}_{y}\right)^{\mathrm{T}}$
can be obtained.

\ELSE

\STATE $\mathbf{n}_{\unit{LED}}^{\textrm{w}}\leftarrow\left(P_{i}^{\textrm{w}}-P_{j}^{\textrm{w}}\right)\times\left(P_{i}^{\textrm{w}}-P_{k}^{\textrm{w}}\right)$,
where $i,j,k\in\left\{ 1,2,3,4\right\} ,i\neq j\neq k$.

\FOR {$i=0\rightarrow\frac{H_{\mathrm{m}}}{\varepsilon_{1}}$}

\STATE $\hat{t}_{z,i}\leftarrow i\varepsilon_{1}$.

\STATE Calculate $\mathbf{\hat{R}}_{\mathrm{c}}^{\mathrm{w}}\left(t_{z,i}\right)$
and $\mathbf{\hat{t}_{\mathrm{c,2D}}^{\mathrm{w}}}\left(t_{z,i}\right)$.

\STATE $\mathbf{\hat{n}}_{\mathrm{LED}}^{\mathrm{w}}\left(t_{z,i}\right)\leftarrow\mathbf{\hat{R}}_{\mathrm{c}}^{\mathrm{w}}\left(t_{z,i}\right)\cdot\mathbf{n}_{\mathrm{LED}}^{\mathrm{c}}$.

\STATE $\Delta G\left(t_{z,i}\right)\leftarrow\left\Vert \mathbf{n}_{\unit{LED}}^{\textrm{w}}-\mathbf{\hat{n}}_{\unit{LED}}^{\textrm{w}}\left(t_{z,i}\right)\right\Vert _{2}$.

\ENDFOR

\STATE $\hat{t}_{z}\leftarrow\min_{t_{z}}\Delta G\left(t_{z}\right)$.
Meanwhile, $\mathbf{\hat{R}}_{\mathrm{c}}^{\mathrm{w}}\leftarrow\mathbf{\hat{R}}_{\mathrm{c}}^{\mathrm{w}}\left(\hat{t}_{z}\right)$
and $\mathbf{\hat{t}_{\mathrm{c,2D}}^{\mathrm{w}}}=\left(\hat{t}_{x},\hat{t}_{y}\right)^{\mathrm{T}}\leftarrow\mathbf{\hat{t}_{\mathrm{c,2D}}^{\mathrm{w}}}\left(\hat{t}_{z}\right)$.

\ENDIF

\ENDIF

\end{algorithmic}

\textbf{Output:} $\mathbf{\hat{R}}_{\mathrm{c,est}}^{\mathrm{w}}$
and $\mathbf{\hat{t}_{\mathrm{c,est}}^{\mathrm{w}}}=\left(\hat{t}_{x},\hat{t}_{y},\hat{t}_{z}\right)^{\mathrm{T}}$.

\label{algorithm1}
\end{algorithm*}

\section{\label{sec:simulation}SIMULATION RESULTS AND ANALYSES}

\global\long\def\arraystretch{0.9}%
\begin{table}[t]
\centering{}\centering{}\caption{\label{tab:Parameters-used-for}System Parameters.}
\begin{tabular}{>{\raggedright}m{5.5cm}|>{\raggedright}m{1.6cm}|>{\centering}m{2.5cm}}
\hline
{\footnotesize{}{}{}{}{}{}{}{}{}Parameter}  & \multicolumn{2}{c}{{\footnotesize{}{}{}{}{}{}{}{}{}Value}}\tabularnewline
\hline
{\footnotesize{}{}{}{}{}{}{}{}{}Room size ($\textrm{length}\times\textrm{width}\times\textrm{height}$)}  & \multicolumn{2}{c}{{\footnotesize{}{}{}{}{}{}{}{}{}$5\,\unit{m}\times5\,\unit{m}\times3\,\unit{m}$}}\tabularnewline
\hline
\multirow{2}{5.5cm}{{\footnotesize{}{}{}{}{}{}{}{}{}Length and width of LED luminaire}} & \centering{}{\footnotesize{}{}{}{}{}{}{}{}{}Length}  & {\footnotesize{}{}{}{}{}{}{}{}{}Width}\tabularnewline
\cline{2-3} \cline{3-3}
 & \multirow{1}{1.6cm}{\centering{}{\footnotesize{}{}{}{}{}{}{}{}{}$120\,\mathrm{cm}$}} & \centering{}{\footnotesize{}{}{}{}{}{}{}{}{}$20\,\mathrm{cm}-100\,\mathrm{cm}$}\tabularnewline
\hline
{\footnotesize{}{}{}{}{}{}{}{}{}LED semi-angle, $\Phi_{\nicefrac{1}{2}}$}  & \multicolumn{2}{c}{{\footnotesize{}{}{}{}{}{}{}{}{}$60{^{\circ}}$}}\tabularnewline
\hline
{\footnotesize{}{}{}{}{}{}{}{}{}Principal point of camera}  & \multicolumn{2}{c}{{\footnotesize{}{}{}{}{}{}{}{}{}$\left(u_{0},v_{0}\right)=\left(320,240\right)$}}\tabularnewline
\hline
{\footnotesize{}{}{}{}{}{}{}{}{}Focal ratio of camera}  & \multicolumn{2}{c}{{\footnotesize{}{}{}{}{}{}{}{}{}$f_{u}=f_{v}=800$}}\tabularnewline
\hline
{\footnotesize{}{}{}{}{}{}{}{}{}The distance between the PD
and the camera in eCA-RSSR, $d_{\mathrm{pc}}$}  & \multicolumn{2}{c}{{\footnotesize{}{}{}{}{}{}{}{}{}$1\,\mathrm{cm}$}}\tabularnewline
\hline
\end{tabular}
\end{table}

Since V-P4L combines VLC and computer vision based localization, a
VLP algorithm named enhanced camera assisted received signal strength
ratio algorithm (eCA-RSSR) \cite{bai2020enhanced}, and a typical
computer vision algorithm termed P4L algorithm \cite{xiaojian2008analytic}
are conducted as the baselines schemes in this section. \vspace{-0.5cm}

\subsection{Simulation Setup}

The system parameters are listed in Table \ref{tab:Parameters-used-for}.
The LED luminaire is deployed in the center of the ceiling. The length
of the luminaire which is along the $x^{\mathrm{w}}$-axis is set
to $120\,\mathrm{cm}$ \cite{Qiu2016Let,philips}, and the widths
of luminaire which is along the $y^{\mathrm{w}}$-axis are varied
according to configurations. We set the rectangular luminaire tilt
with various angles along the $y^{\mathrm{w}}$-axis to represent
that LEDs have different heights. All statistical results are averaged
over $1000$ independent runs. For each simulation run, the receiver
positions are selected in the room randomly. The pinhole camera is
calibrated. The image noise is modeled as a white Gaussian noise having
an expectation of zero and a standard deviation of $2\;\unit{pixels}$
\cite{zhou2019robust}. Since the image noise affects the pixel coordinate
of the luminaire's projection on the image plane, the pixel coordinate
is obtained by processing 20 images for the same position. Moreover,
we use two-step segmentation optimization strategy, and we set $\varepsilon_{1}=10\:\mathrm{cm}$
and $\varepsilon_{2}=1\:\mathrm{cm}$.

Note that since eCA-RSSR requires three LEDs for localization, we
assume that the four LEDs at the vertices of the luminaire are used
for eCA-RSSR and we choose the three LEDs with the highest RSSs to
achieve best performance for eCA-RSSR. Additionally, all the LEDs
transmit different information in eCA-RSSR, which is not required
in V-P4L and the P4L algorithm. Therefore, compared with V-P4L, the
VLC link of eCA-RSSR is more complex. Furthermore, eCA-RSSR relies
on the perfect Lambertian pattern model. However, the VLC channel
model can be quiet different from the Lambertian pattern model even
using the LED having nearly-ideal Lambertian pattern, as shown in
\cite{miramirkhani2015channel}, and the difference can be over 100\%
in certain cases. Therefore, we set a random deviation $\delta_{\mathrm{1}}\leq10\%$
for the Lambertian pattern model for eCA-RSSR, conservatively. On
the other hand, the P4L algorithm exploits a rectangle to estimate
the position and pose of the camera. The P4L algorithm assumes that
the camera knows the 3D-2D correspondences. However, the beacon in
the P4L algorithm cannot convey time-domain information to the camera,
which make the assumption impractical. In addition, the method to
find the 3D-2D line correspondences given by \cite{ICRA2018Manhattan}
is also not practical when the camera captures the beacons on the
ceiling as stated in Section \ref{sec:intro}. Therefore, we set a
random error rate $\delta_{\mathrm{2}}\leq10\%$ for the 3D-2D correspondences,
conservatively. Moreover, the P4L algorithm can only obtain the relative
position. For comparison with V-P4L, we transform the relative position
into the absolute position for the P4L algorithm in this section.

We evaluate the performance of V-P4L in terms of its accuracy of position
and pose estimation. We define position error as:
\begin{equation}
PE=\left\Vert \mathbf{r}_{\unit{true}}^{\textrm{w}}-\mathbf{r}_{\unit{est}}^{\textrm{w}}\right\Vert ,\label{eq:48}
\end{equation}
where $\mathbf{r}_{\unit{true}}^{\textrm{w}}=\left(x_{r,\unit{true}}^{\textrm{w}},y_{r,\unit{true}}^{\textrm{w}},z_{r,\unit{true}}^{\textrm{w}}\right)$
and $\mathbf{r_{\textrm{est}}^{\textrm{w}}}=\left(x_{r,\mathscr{\textrm{est}}}^{\textrm{w}},y_{r,\textrm{est}}^{\textrm{w}},z_{r,\textrm{est}}^{\textrm{w}}\right)$
are the actual and estimated world coordinates of the receiver, respectively.
Additionally, the accuracy of pose estimation can be measured by the
orientation error which is defined as:
\begin{equation}
OE=\left|\varTheta_{\mathrm{true}}-\varTheta_{\mathrm{est}}\right|,\label{eq:49}
\end{equation}
where $\varTheta_{\mathrm{true}}$ and $\varTheta_{\mathrm{est}}$
are the actual and estimated rotation angles, respectively.

In this section, we will evaluate the performance of V-P4L under various
tilted angles of the luminaire, various widths of the luminaire and
various image noise. We compare for both scenarios where LEDs have
the same height and have different heights, which are denoted by SH
and DH, respectively in the figures. Since the basic algorithm of
V-P4L is used when LEDs have the same height, we denote the basic
algorithm of V-P4L by V-P4L-SH in figures.\vspace{-0.4cm}

\subsection{\label{subsec:Accuracy-Performance-1}Effect Of Luminaire's Tilted
Angle On Accuracy Performance}

\begin{figure*}[t]
\setlength{\abovecaptionskip}{0.2cm} %调整图片标题与图距离
\setlength{\belowcaptionskip}{-8pt} %调整图片标题与下文距离
 \centering \subfigure[Without occlusion.]{\label{secfig1} \includegraphics[scale=0.5]{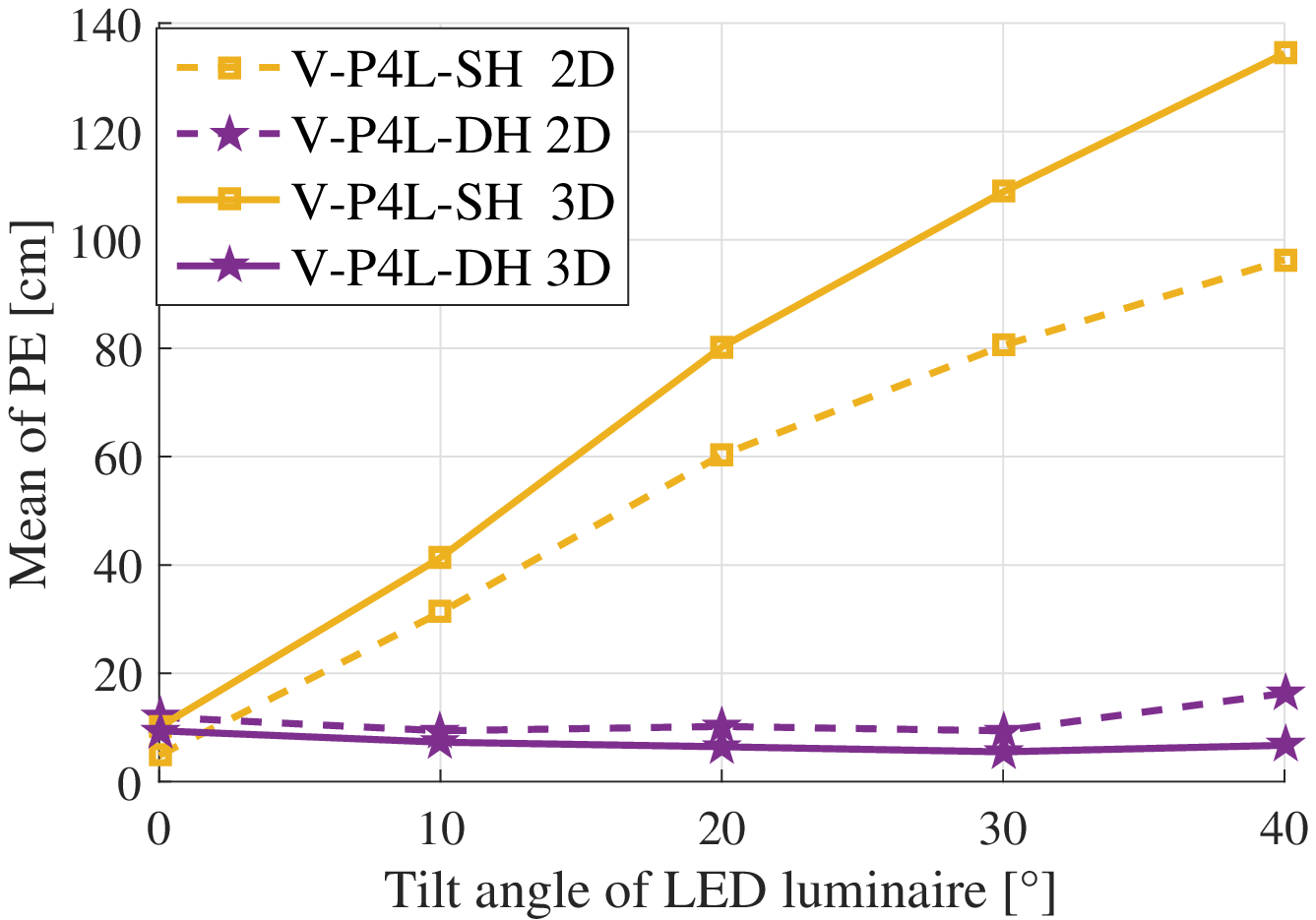}
} \subfigure[With occlusion.]{\label{secfig2} \includegraphics[scale=0.5]{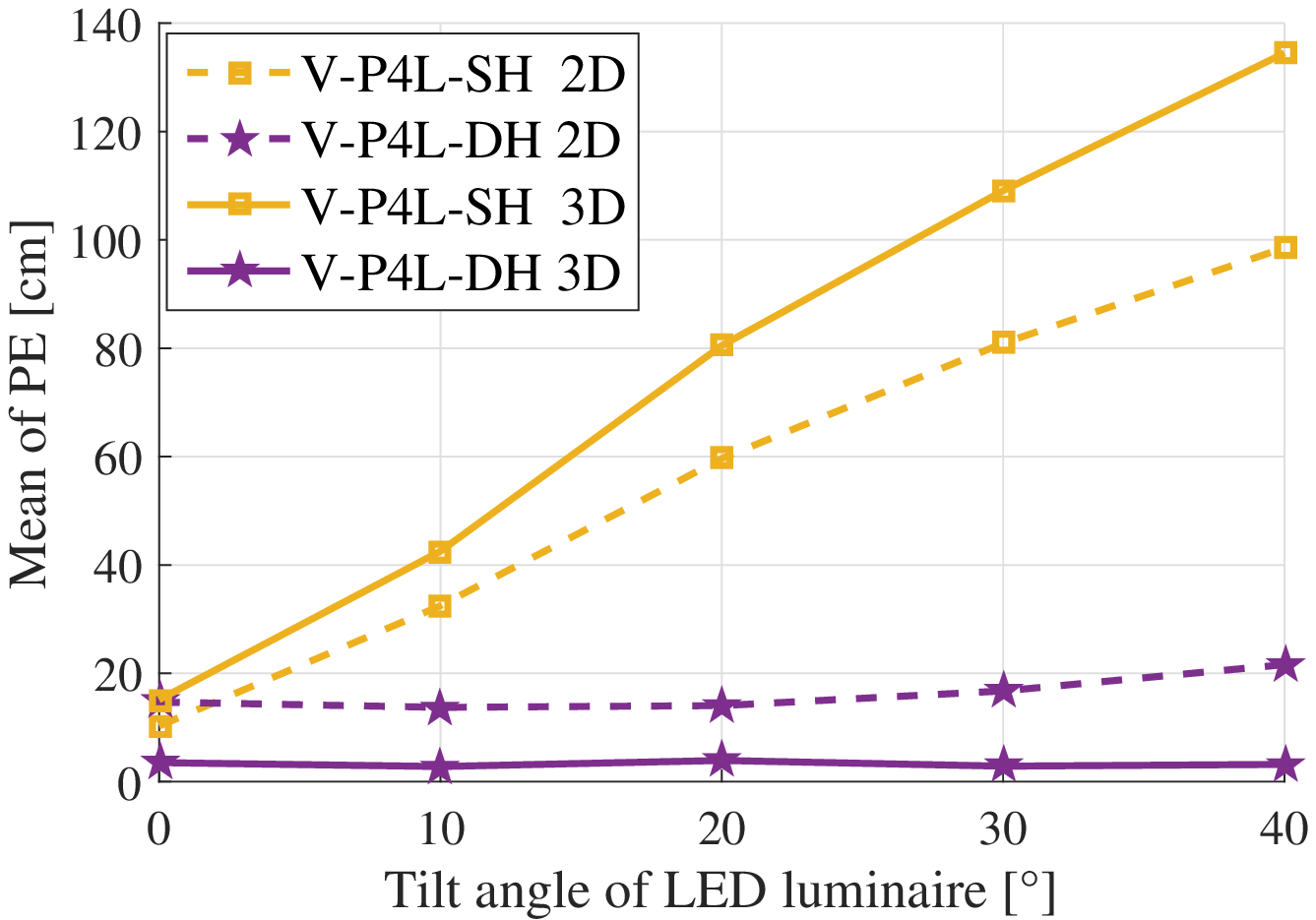}
} \caption{\label{fig:tilt angle PE}The comparison of position errors (PEs)
with varying tilted angles of the luminaire between the basic algorithm
of V-P4L and V-P4L-DH. (a) There is no occlusion. (b) The projection
of $P_{2}$, $p_{2}$, is blocked by barriers and not on the image
plane as shown in Fig. \ref{fig:occlusion}.}
\end{figure*}

\begin{figure*}[t]
\setlength{\abovecaptionskip}{0.2cm} %调整图片标题与图距离
\setlength{\belowcaptionskip}{-8pt} %调整图片标题与下文距离
 \centering \subfigure[OEs along the x-axis.]{ \includegraphics[width=0.31\linewidth]{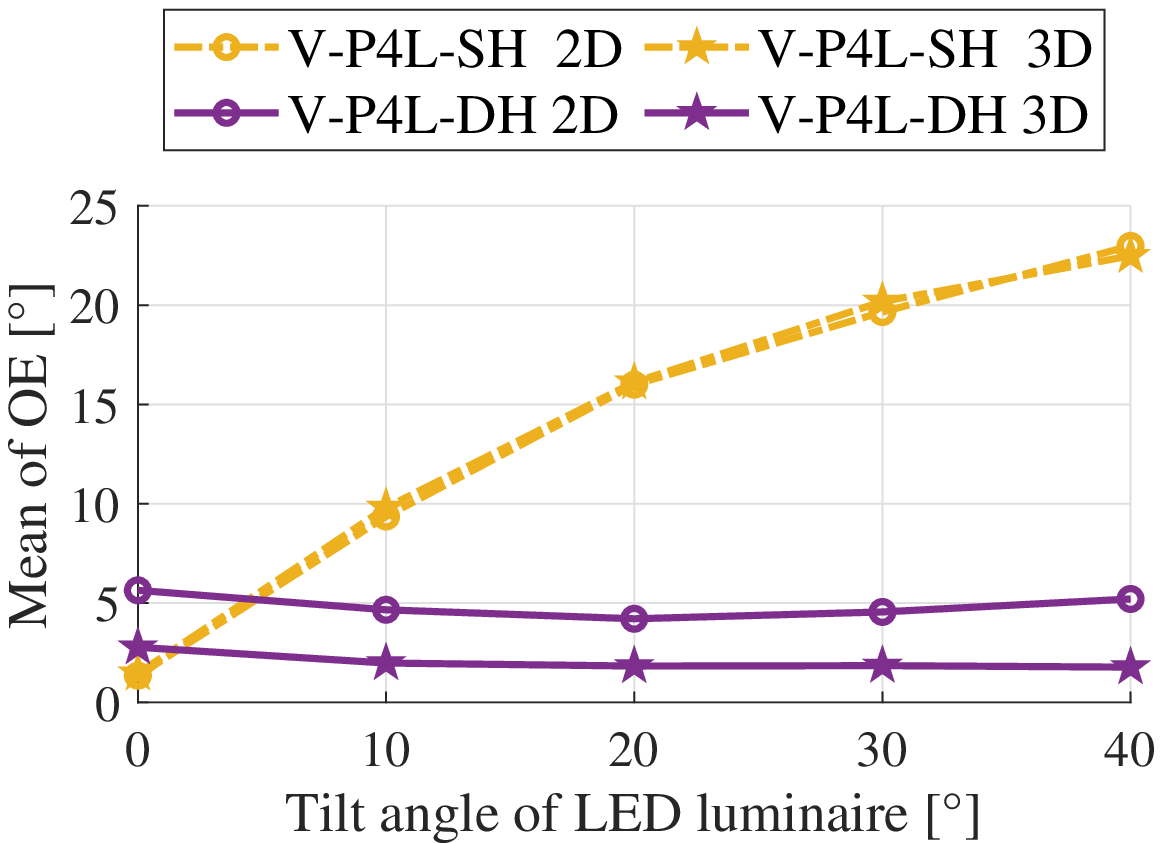}
} \subfigure[OEs along the y-axis.]{ \includegraphics[width=0.31\linewidth]{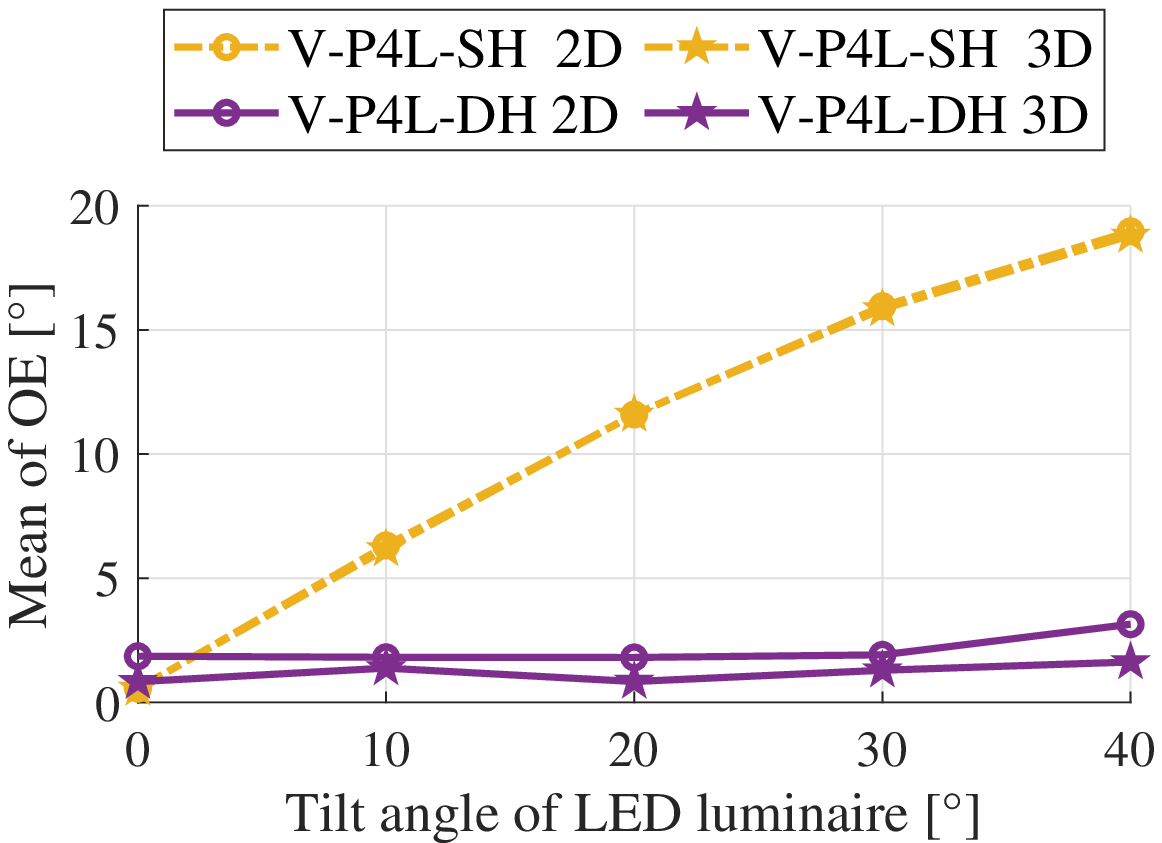}
} \subfigure[OEs along the z-axis.]{ \includegraphics[width=0.31\linewidth]{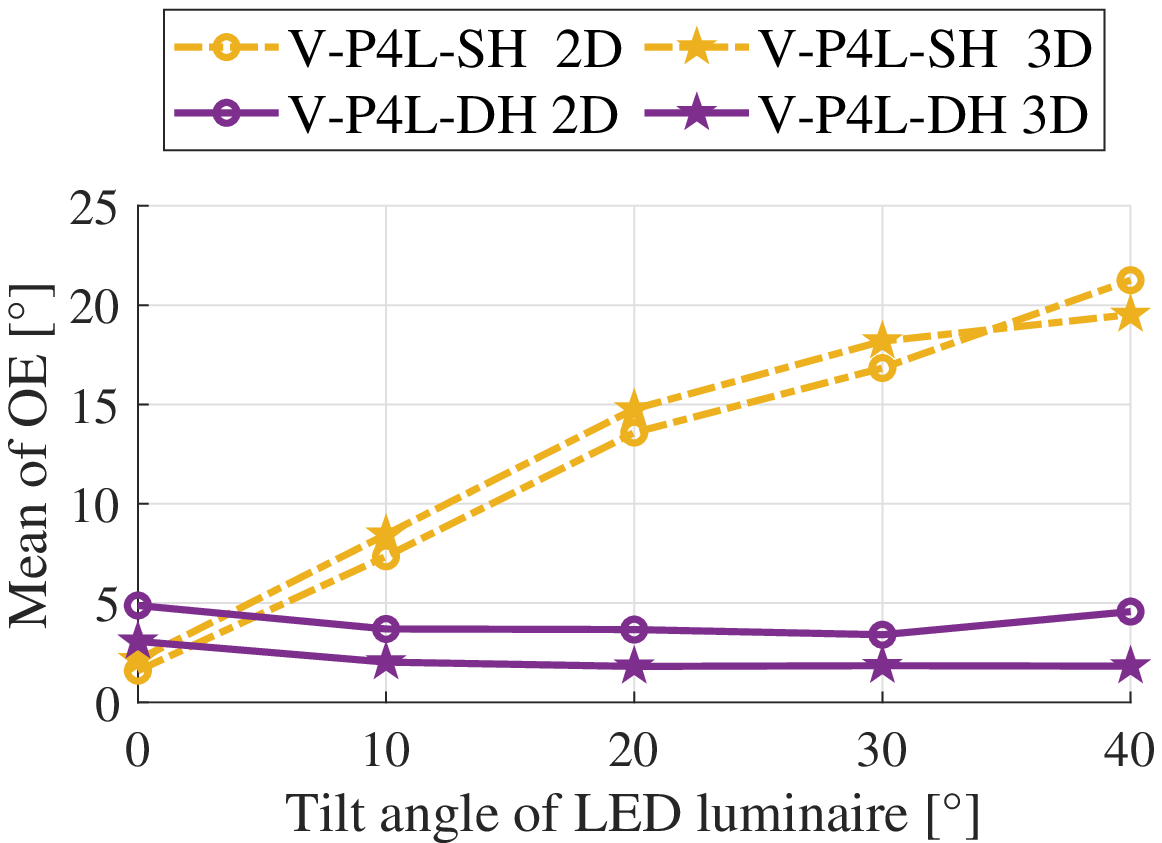}
} \caption{\label{fig:tilt angle OE}The comparison of orientation errors (OEs)
along the $x$-axis, $y$-axis and $z$-axis between V-P4L-SH and
V-P4L-DH.}
\end{figure*}

We first evaluate the effect of the tilted angles of the luminaire
on localization accuracy of V-P4L for both 2D and 3D localization.
This performance is represented by the means of PEs with the tilted
angles of the luminaire varying from $0{^{\circ}}$ to $40{^{\circ}}$.
The width of the luminaire is $40\;\unit{cm}$. As shown in Fig. \ref{secfig1},
for 2D localization, when the tilted angle of the luminaire is $0{^{\circ}}$,
i.e., LEDs have the same height, the basic algorithm of V-P4L can
obtain a slight better performance than V-P4L-DH. However, as the
tilted angle of the luminaire varying from $0{^{\circ}}$ to $40{^{\circ}}$,
the means of PEs of the basic algorithm of V-P4L increase from about
5 cm to about 98 cm. For 3D localization, the means of PE of the basic
algorithm of V-P4L increase from about 10 cm to about 135 cm. In contrast,
for all the tilted angles of the luminaire, the means of PEs of V-P4L-DH
are less than 18 cm for both 2D and 3D localization. Since in 3D localization
the estimated 2D position and pose is optimized meanwhile when search
for the optimal $\hat{t}_{z}$, V-P4L can achieve better performance
for 3D localization than 2D localization.

Since the accuracy of pose estimation is also affected by the tilted
angle of the luminaire, we then evaluated the effect of the tilted
angles on OEs of V-P4L for both 2D and 3D localization. As shown in
Fig. \ref{fig:tilt angle OE}, the OEs along the $x$-axis, $y$-axis
and $z$-axis are shown seperately. When the tilted angle of the luminaire
is $0{^{\circ}}$, the basic algorithm of V-P4L can obtain slight
better performance than V-P4L-DH. However, as the tilted angle of
the luminaire varying from $0{^{\circ}}$ to $40{^{\circ}}$, the
means of OEs of the basic algorithm of V-P4L increase from about $2{^{\circ}}$
to about $20{^{\circ}}$. In contrast, for all the tilted angles of
the luminaire, V-P4L-DH can achieve consistent well performance, and
the means of OEs are always less than $6{^{\circ}}$.

We also evaluate the effect of partial occlusion as shown in Fig.
\ref{fig:occlusion} on localization accuracy of V-P4L. Figure \ref{secfig1}
shows the accuracy performance of position estimation when there is
no occlusion in the scenario. In contrast, Fig. \ref{secfig2} shows
the accuracy performance of position estimation when $p_{2}$ is blocked
and is not on the image plane. As shown in Fig. \ref{secfig2}, when
the luminaire is partially blocked, V-P4L can still achieve high accuracy.
In particular, the accuracy performance of the basic algorithm of
V-P4L is almost the same in Fig. \ref{secfig1} and Fig. \ref{secfig2}.
In addition, for V-P4L-DH, the 2D-localization accuracy reduces about
2 cm, while the 3D-localization accuracy improves about 4 cm. Therefore,
V-P4L is robust to partial occlusion as introduced in Section \ref{sec:V-P4L-DH}.

In this subsection, we have verified V-P4L can achieve high accuracy
for both 2D and 3D localization. Since 2D localization is the special
case of 3D localization where the height of the receiver is known
in advance, in the following subsections, we will only show the simulation
results for 3D localization.

\subsection{\label{subsec:Accuracy-Performance}Effect Of Luminaire's Widths
On Accuracy Performance}

\begin{figure}
\begin{centering}
\includegraphics[scale=0.5]{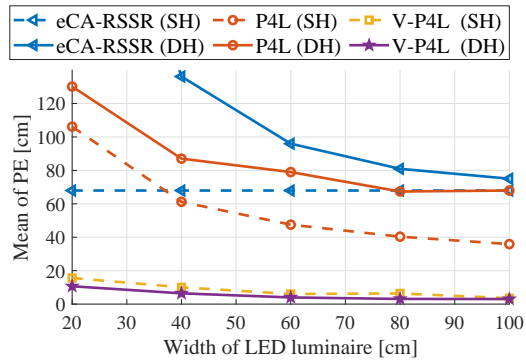}
\par\end{centering}
\caption{\label{fig:width compare1}The comparison of position errors (PEs)
with varying widths of the luminaire among eCA-RSSR, the P4L algorithm
and V-P4L for both the scenarios where LEDs have the same height and
have different heights.}
\end{figure}

We then evaluate the effect of the luminaire's width on localization
accuracy of V-P4L. This performance is represented by the means of
PEs with the width varying from $20\;\unit{cm}$ to $100\;\unit{cm}$.
For the scenarios where LEDs have different heights, the tilted angle
of the luminaire is $20{^{\circ}}$. As shown in Fig. \ref{fig:width compare1},
for both the scenarios where LEDs have the same height and have different
heights, V-P4L is able to obtain the best performance among the three
algorithms. When LEDs have the same height, the means of PEs of the
basic algorithm of V-P4L are below $15\;\unit{cm}$. In contrast,
for eCA-RSSR, the means of PEs are around 70 cm as the width of the
luminaire increases from $20\;\unit{cm}$ to $100\;\unit{cm}$. Additionally,
for the P4L algorithm, the means of PEs decrease from over $100\;\unit{cm}$
to about $40\;\unit{cm}$. On the other hand, when LEDs have different
heights, the means of PEs of V-P4L-DH decrease from about 10 cm to
5 cm. In contrast, the means of PEs of both eCA-RSSR and the P4L algorithm
are higher than 50 cm for all the widths of the luminaire. As shown
in Fig. \ref{fig:width compare1}, the localization accuracy increases
with the increase of the width for V-P4L. However, the PEs of V-P4L
are always less than $15\,\unit{cm}$ regardless of the height differences
among LEDs using a single LED luminaire whose width is longer than
$20\;\unit{cm}$, and thus V-P4L can be applied to popular indoor
luminaires.

\begin{figure*}[t]
\setlength{\abovecaptionskip}{0.2cm} %调整图片标题与图距离
\setlength{\belowcaptionskip}{-8pt} %调整图片标题与下文距离
 \centering \subfigure[OEs along the x-axis.]{ \includegraphics[width=0.31\linewidth]{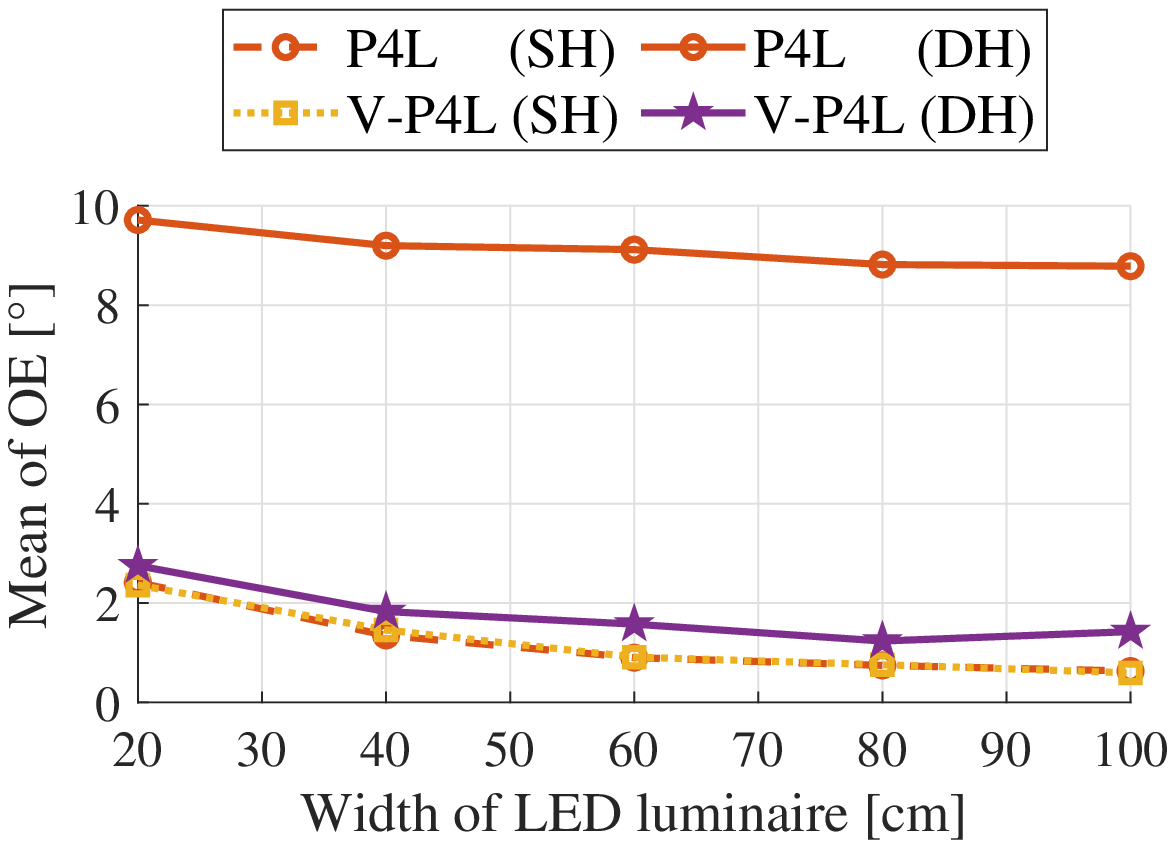}
} \subfigure[OEs along the y-axis.]{ \includegraphics[width=0.31\linewidth]{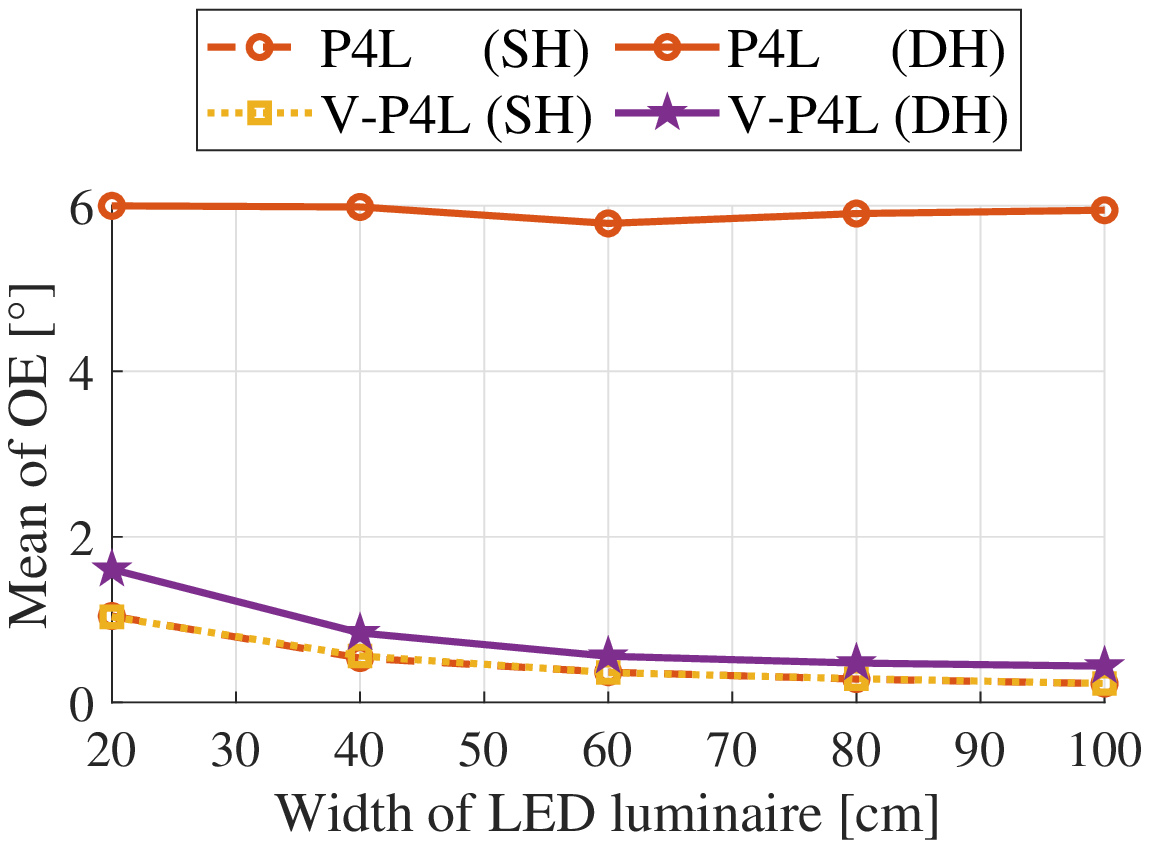}
} \subfigure[OEs along the z-axis.]{ \includegraphics[width=0.31\linewidth]{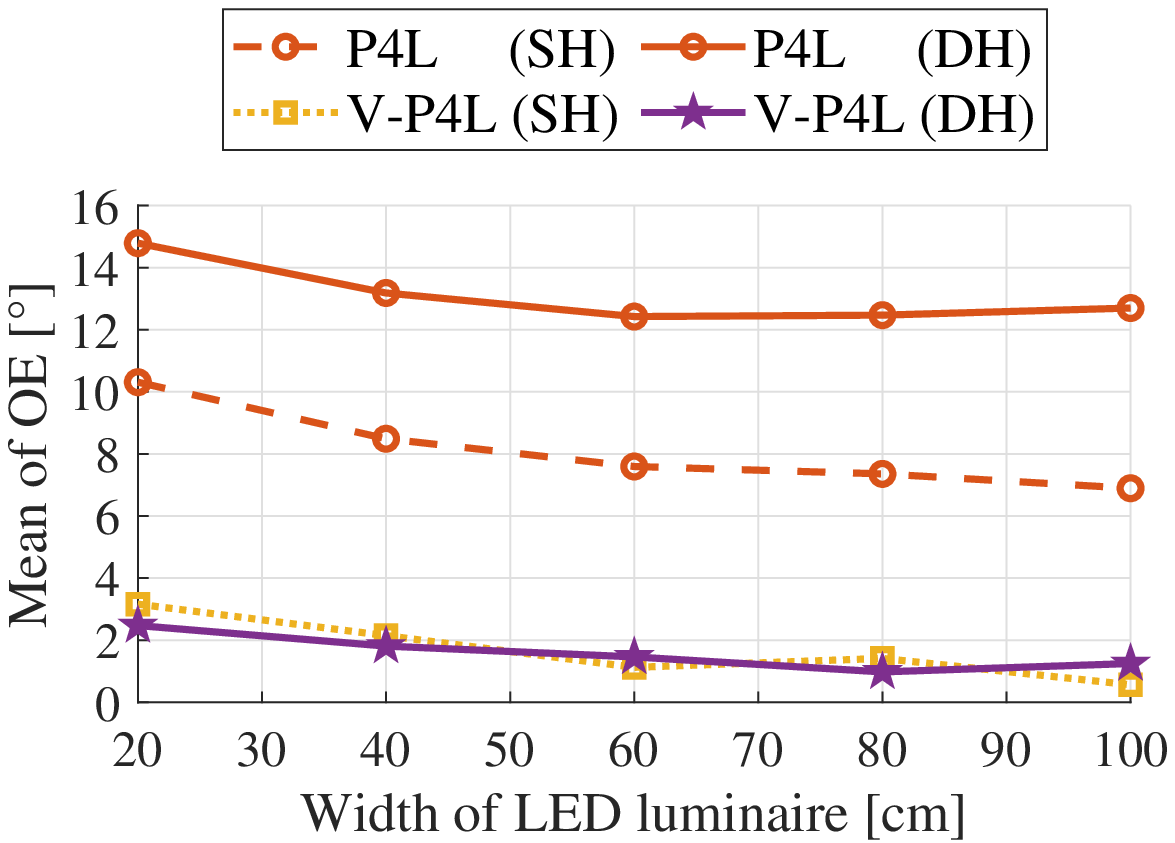}
} \caption{\label{fig:width orientation 3D}The comparison of orientation errors
(OEs) along the $x$-axis, $y$-axis and $z$-axis between the P4L
algorithm and V-P4L with varying widths of the luminaire.}
\end{figure*}

Since the accuracy of pose estimation is also affected by the width
of the LED luminaire, we then compare the OEs between V-P4L and the
P4L algorithm with varying widths of the luminaire. As shown in Fig.
\ref{fig:width orientation 3D}, for V-P4L, all the means of OEs along
the $x$-axis, $y$-axis and $z$-axis are less than $3.5{^{\circ}}$
regardless of the height differences among the LEDs. When LEDs have
the same height, the means of OEs along the $x$-axis and $y$-axis
decrease from about $2{^{\circ}}$ to about $0.5{^{\circ}}$ and from
$1{^{\circ}}$ to about $0.3{^{\circ}}$, respectively for both the
basic algorithm of V-P4L and the P4L algorithm. Additionally, the
means of OEs of the basic algorithm of V-P4L along the $z$-axis decrease
from about $3{^{\circ}}$ to about $0.5{^{\circ}}$, which is over
$5{^{\circ}}$ better than that of the P4L algorithm. On the other
hand, when LEDs have different heights, for V-P4L-DH, the means of
OEs along the $x$-axis, $y$-axis and $z$-axis decrease from $2.5{^{\circ}}$
to $1.5{^{\circ}}$, from $1.5{^{\circ}}$ to $0.5{^{\circ}}$ and
from $2.2{^{\circ}}$ to $1.5{^{\circ}}$, respectively. In contrast,
for the P4L algorithm, the means of OEs along the $x$-axis, $y$-axis
and $z$-axis are about $9{^{\circ}}$, $6{^{\circ}}$, $13{^{\circ}}$,
respectively. Therefore, compared with the P4L algorithm, V-P4L can
obtain higher accuracy for pose estimation using popular indoor luminaire.

\subsection{\label{subsec:noise effect}Effect Of Image Noise On Accuracy Performance}

\begin{figure}
\begin{centering}
\includegraphics[scale=0.5]{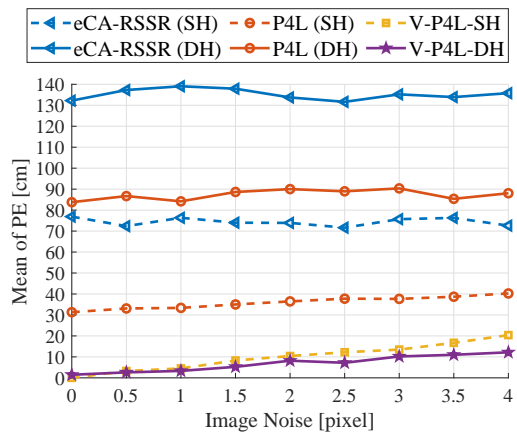}
\par\end{centering}
\caption{\label{fig:noise PE}The comparison of position errors (PEs) with
varying image noise among eCA-RSSR, the P4L algorithm, V-P4L for both
the scenarios where LEDs are at the same height and at different heights.}

\end{figure}

In this subsection, we evaluate the effect of the image noise on the
localization performance of V-P4L when the width of the luminaire
is $40\;\unit{cm}$. When LEDs have different heights, the tilted
angle of the luminaire is $20{^{\circ}}$. The image noise is modeled
as a white Gaussian noise having an expectation of zero and a standard
deviation, $\sigma_{n}$, ranging from $0$ to $4\;\unit{pixels}$
\cite{zhou2019robust}. Figure \ref{fig:noise PE} shows the means
of PEs versus image noises. As shown in Fig. \ref{fig:noise PE},
when LEDs have the same height, the means of PEs of the basic algorithm
of V-P4L increase from $0\;\unit{cm}$ to $20\;\unit{cm}$ as the
image noise increases from $0$ to $4\;\unit{pixels}$. In contrast,
for the P4L algorithm, the means of PEs increase from 30 cm to 40
cm. Additionally, for eCA-RSSR, the means of PEs are around 75 cm.
When LEDs have different heights, the means of PEs of V-P4L-DH increase
from 0 cm to 11 cm. In contrast, for the P4L algorithm, the means
of PEs are about 85 cm. Additionally, for eCA-RSSR, the means of PEs
are about 135 cm. Therefore, compared with the P4L algorithm and eCA-RSSR,
V-P4L can obtain higher accuracy for position estimation.

\begin{figure*}[t]
\setlength{\abovecaptionskip}{0.2cm} %调整图片标题与图距离
\setlength{\belowcaptionskip}{-8pt} %调整图片标题与下文距离
 \centering \subfigure[OEs along the x-axis.]{ \includegraphics[width=0.31\linewidth]{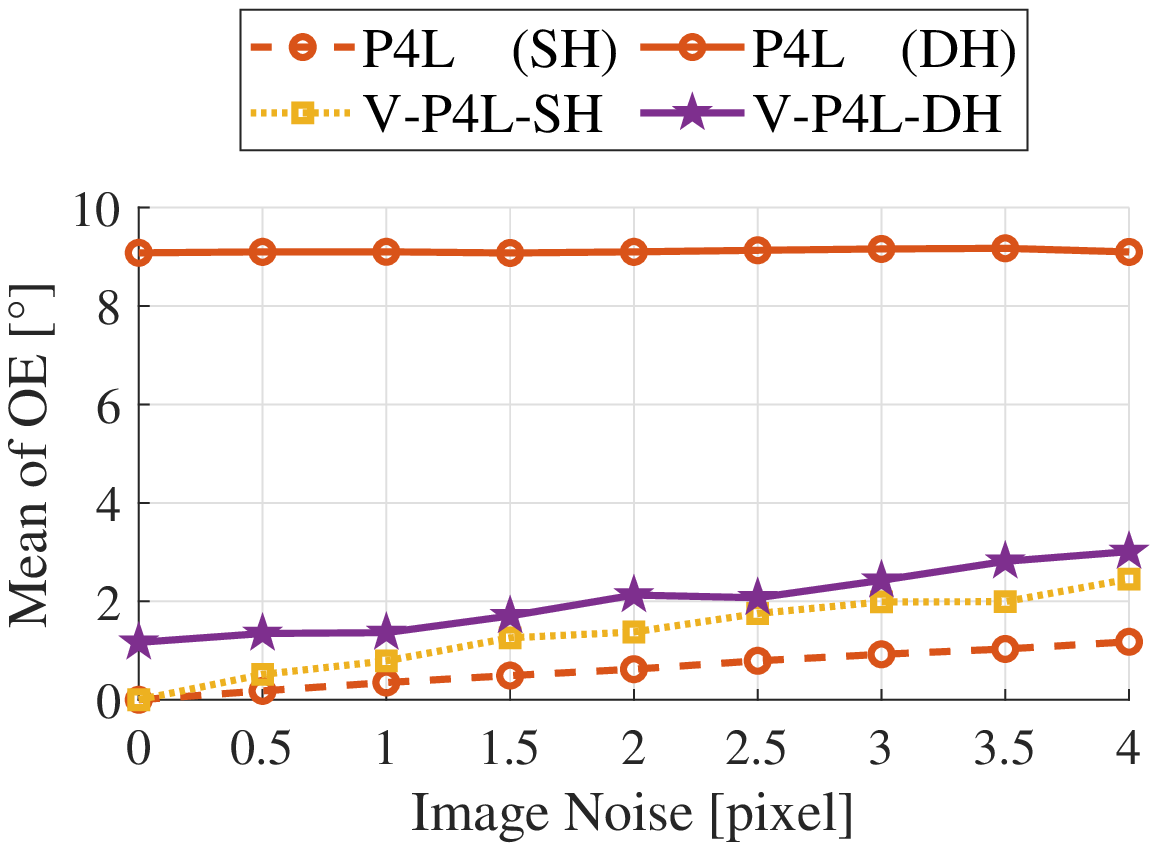}
} \subfigure[OEs along the y-axis.]{ \includegraphics[width=0.31\linewidth]{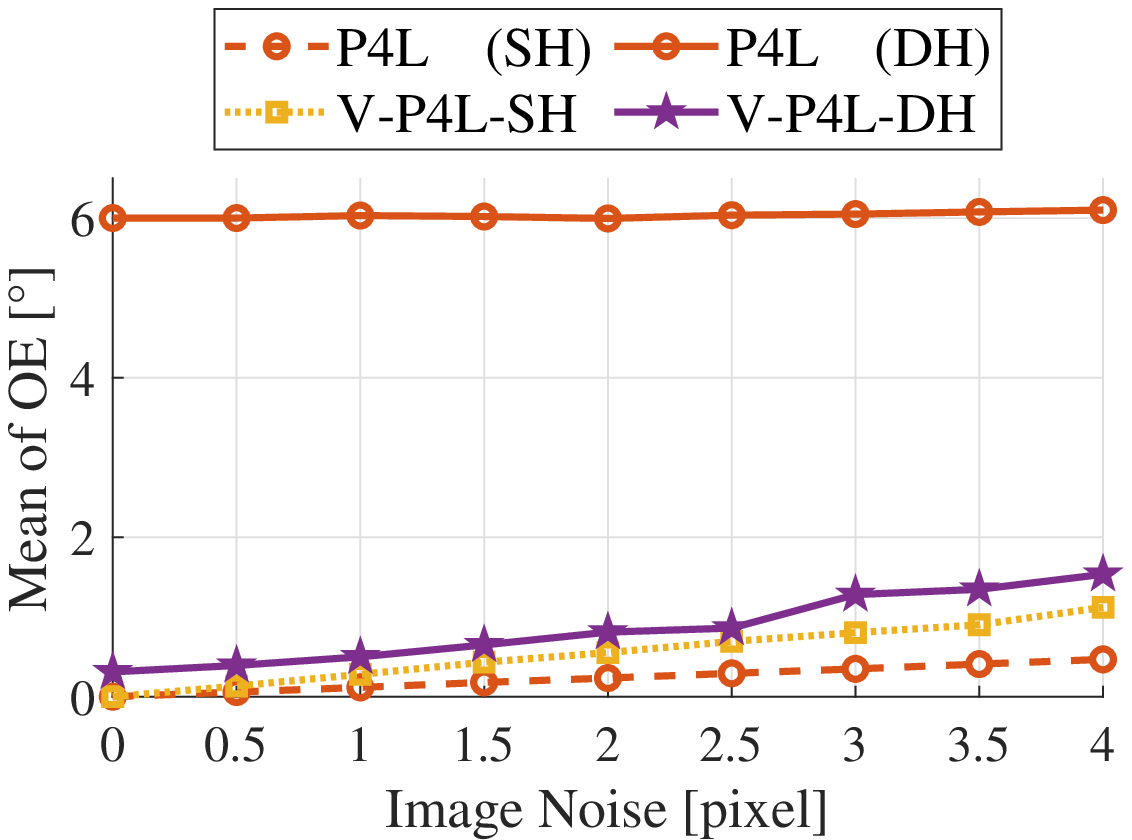}
} \subfigure[OEs along the z-axis.]{ \includegraphics[width=0.31\linewidth]{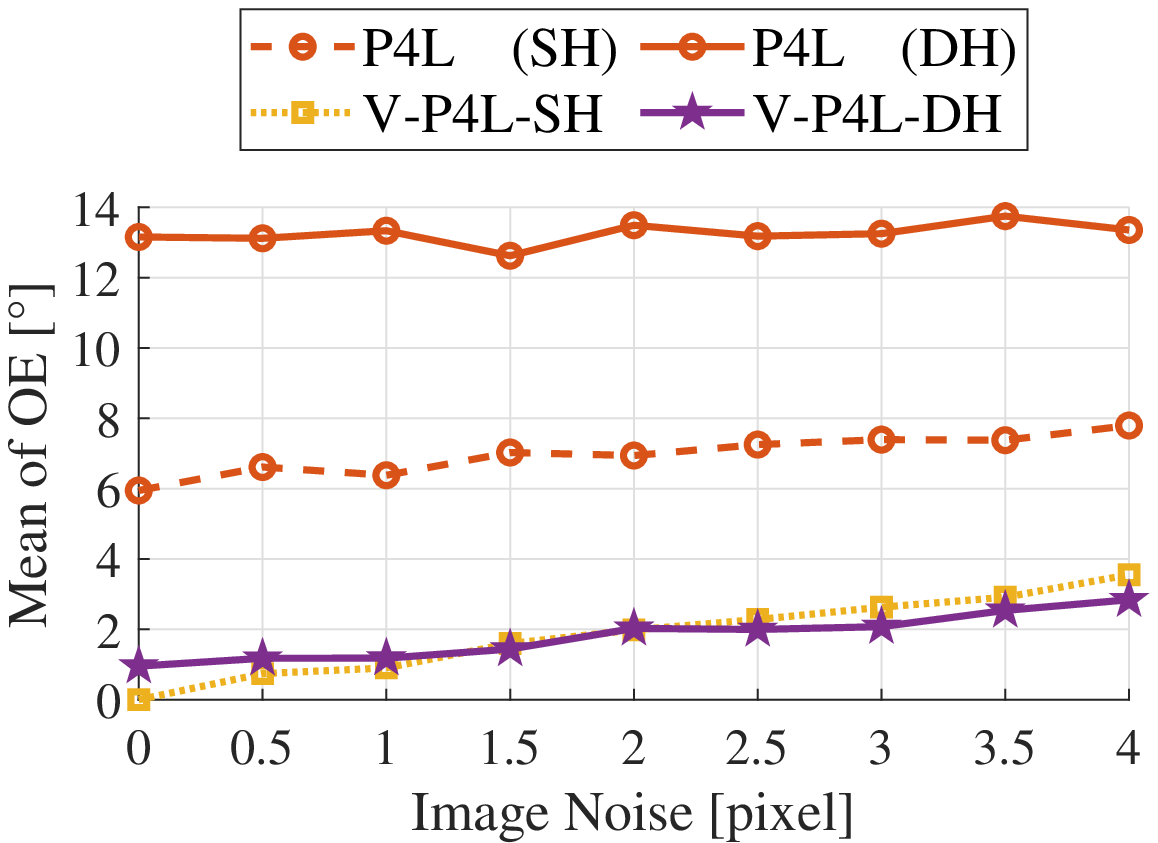}
} \caption{\label{fig:noise OE 3D}The comparison of orientation errors (OEs)
along the $x$-axis, $y$-axis and $z$-axis between the P4L algorithm
and V-P4L with varying image noise.}
\end{figure*}

Finally, we compare the accuracy of pose estimation between V-P4L
and the P4L algorithm under different image noises. Figure \ref{fig:noise OE 3D}
show the means of orientation errors along $x$-axis, $y$-axis and
$z$-axis with the image noise ranging from $0$ to $4\;\unit{pixels}$.
When LEDs have the same height, the means of OEs of the basic algorithm
of V-P4L along $x$-axis, $y$-axis and $z$-axis increase from $0{^{\circ}}$
to $2.5{^{\circ}}$, from $0{^{\circ}}$ to $1.2{^{\circ}}$ and from
$0{^{\circ}}$ to $3.8{^{\circ}}$, respectively. In contrast, for
the P4L algorithm, the means of OEs along the $x$-axis and $y$-axis
increase from $0{^{\circ}}$ to $1.2{^{\circ}}$ and from $0{^{\circ}}$
to $0.5{^{\circ}}$, respectively, which is slight better than the
basic algorithm of V-P4L. However, the means of OEs of the P4L algorithm
along the $z$-axis increase from about $6{^{\circ}}$ to about $8{^{\circ}}$
which is over $4{^{\circ}}$ worse than that of the basic algorithm
of V-P4L. On the other hand, when LEDs have different heights, for
V-P4L-DH, the means of OEs along $x$-axis, $y$-axis and $z$-axis
increase from about $1{^{\circ}}$ to less than $3.5{^{\circ}}$.
In contrast, for the P4L algorithm, the means of OEs along $x$-axis,
$y$-axis and $z$-axis are about $9{^{\circ}}$, $6{^{\circ}}$,
$13{^{\circ}}$, respectively. Therefore, compared with the P4L algorithm,
V-P4L can obtain more stable and accurate pose estimation regardless
of the image noise.

\vspace{-0.3cm}

\section{CONCLUSION}

\label{sec:CONCLUSION}We have proposed a novel indoor localization
algorithm named V-P4L that estimates the position and pose of the
camera using a single, VLC-enabled LED luminaire. The camera is used
to simultaneously capture the information in both time and space domains.
Based on the information captured by the camera, V-P4L does not require
the 3D-2D correspondences. Moreover, V-P4L can be implemented regardless
of the height differences among LEDs. Therefore, V-P4L can achieve
higher feasibility and higher accuracy than eCA-RSSR and the conventional
PnL algorithms. Simulation results have shown that for V-P4L the position
error is always less than $15\,\unit{cm}$ and the orientation error
is always less than $3{^{\circ}}$ using popular indoor luminaires.
In the future, we will experimentally implement V-P4L based on a dedicated
test bed.

\vspace{-0.6cm}

% -------------------------------------------------------------------------
 \bibliographystyle{IEEEbib}
\bibliography{BL_abbr_all}

\end{document}